\documentclass[12pt]{article}

\usepackage[english]{babel}
\usepackage{lmodern}
\usepackage{xcolor}
\usepackage{array}
\usepackage{amsmath}
\usepackage{amssymb}
\usepackage{amsfonts}
\usepackage{amsthm}
\usepackage{enumitem}
\usepackage{mathabx}
\usepackage{pgf,tikz}
\usepackage{bm}
\usepackage{natbib}
\usepackage{authblk}
\usepackage{subcaption}
\usepackage[bookmarks={true},bookmarksopen={true}]{hyperref}
\usepackage{algorithmic, algorithm}
\usepackage{multirow}
\usepackage{booktabs}

\setcitestyle{numbers,square} 

\newtheorem{prop}{Proposition}[section]

\newtheorem{rem}[prop]{Remark}

\newcommand{\R}{\mathbb{R}}
\newcommand{\Z}{\mathbb{Z}}

\newcommand{\C}{\mathbb{C}}

\newcommand{\B}{{\mathbf B}}

\newcommand{\der}{\mathrm{d}}

\newcommand{\E}{\mathbb{E}}

\renewcommand{\epsilon}{\varepsilon}
\newcommand{\1}{\mathbf{1}}
\newcommand{\bv}{\bm v}
\newcommand{\be}{\bm e}
\newcommand{\bw}{\bm w}

\graphicspath{{.}{./}{./manuscript_files/}}

\title{On simulation of continuous determinantal point processes}
\author[1,2]{Fr\'ed\'eric Lavancier}
\author[3]{Ege Rubak}
\affil[1]{Laboratoire de Math\'ematiques Jean Leray, BP 92208,  2 Rue de la Houssini\`ere, F-44322 Nantes Cedex 03, France, \texttt{Frederic.Lavancier@univ-nantes.fr}.}
\affil[2]{CREST-ENSAI, UMR CNRS 9194, Campus de Ker-Lann, rue Blaise Pascal, BP 37203, 35172 Bruz cedex, France, \texttt{Frederic.Lavancier@ensai.fr}.}
\affil[3]{Department of Mathematical Sciences\\
Aalborg University, \texttt{rubak@math.aau.dk}.}

\begin{document}
\maketitle

\begin{abstract}
  We review how to simulate continuous determinantal point processes (DPPs) and improve the current simulation
  algorithms in several important special cases as well as detail how certain types of conditional
  simulation can be carried out. Importantly we show how to speed up the simulation of the widely
  used Fourier based projection DPPs, which arise as approximations of more general DPPs. 
  The
  algorithms are implemented and published as open source software.
  
  \medskip
  

\end{abstract}

\section{Introduction}

Since the seminal work by A.\ Kulesza and B.\ Taskar \cite{Kulesza:Taskar:12}, discrete DPPs defined on a finite set $\{1,\dots,N\}$ have become popular objects in the machine learning community to generate diverse random subsets, the main application being for recommandation systems.
Discrete DPPs are defined through a kernel matrix of size $N\times N$, which is often assumed to be Hermitian.
The default algorithm to make perfect simulation of discrete DPPs, the so-called spectral algorithm of \cite{HKPV06}, starts from an orthonormal eigendecomposition of this matrix.
Since this decomposition may be costly to obtain, and even unfeasible for large $N$, other options have been developed, from approximate algorithms \cite{affandi13a,Li16} to alternative perfect simulation strategies \cite{gillenwater19a,Derezinski19,poulson2020,launay20}. 

On the other side, continuous DPPs generally refer to DPPs defined on the Euclidean space $\R^d$.
They constitute the initial setting of O.\ Macchi \cite{macchi1975} who introduced these processes in their current form to model the distribution of a fermion system in statistical physics.
Continuous DPPs find applications in machine learning \cite{affandi2013approximate}, spatial statistics \cite{Lavancier}, telecommunications \cite{deng2014, miyoshi2014} and Monte Carlo approximations \cite{bardenet2020, belhadji2019kernel,coeurjolly2020monte}. 
They  are defined through a kernel $K$ which in this setting is a function $K(x,y)$, $x,y\in \R^d$,
that must satisfy some properties to ensure existence of the continuous DPP as detailed in Section~\ref{preliminaries}. 
The spectral algorithm of \cite{HKPV06} to simulate continuous DPPs on a compact set $S\subset\R^d$ is in general feasible only if we know the spectral representation (also called Mercer representation) of $K$ on $S$. While this representation always exists in theory, it is in general intractable. 
Unfortunately, the aforementioned alternative perfect simulation strategies developed in the discrete case do not apply to the continuous case, and their extension to this setting is a difficult open challenge. 
For these reasons, there is still an avenue to improve simulation of continuous DPPs.
This paper addresses this question for projection DPPs and other important classes of DPPs, including the $\beta$-Ginibre process, the Gaussian-type DPP and the Bessel-type DPP. 
 
A projection DPP on a bounded subset $S\subset\R^d$ corresponds to the special case where  $K$ satisfies the identity $\int_S K(x,z)K(z,y)dz= K(x,y)$.
This class of DPPs is of particular interest for several reasons.
First, unlike general DPPs, projection DPPs generate a fixed number of points in $S$, equal to $n=\int_S K(x,x)dx$. 
Second, they correspond to the most repulsive DPPs, since their (non null) eigenvalues in their spectral representation are all identically equal to 1, the maximal possible value.
Third, the simulation algorithm of \cite{HKPV06} for a  general DPP, provided its spectral representation is known,  boils down to the simulation of an associated projection DPP.
A first contribution of this paper, carried out in Section~\ref{sec:projection},  is to review and improve the simulation algorithms of projection DPPs.
An important particularity of these processes is that they can be perfectly simulated without
knowing the spectral representation of their kernel; see Algorithm~\ref{alg:nospectral}. 
Nonetheless, if we have access to this   representation, then the well-known spectral algorithm (Algorithm~\ref{alg:sim}) is more efficient.
As a special case, we give a particular account of the case when the eigenfunctions of the spectral
decomposition are the Fourier basis functions on $S$.
This case plays an important role both because it gives rise to an homogeneous projection DPP on
$S$ and because it is at the heart of the spectral approximation of general invariant-translation kernels in \cite{Lavancier}.
For this basis, we refine the spectral algorithm to decrease the computation time by up to $45\%$
in dimension $d=1$ and up to $25\%$ in $d=2$; see Algorithm~\ref{alg:rejection}.
Finally, we show how  Algorithm~\ref{alg:nospectral} and  Algorithm~\ref{alg:sim} can be combined to perform conditional simulation of a projection DPP given a subset of points on $S$,  including in-painting conditional simulation.

For general (non projection) DPPs, the standard simulation procedure consists in approximating the spectral representation of their kernel in order to apply the algorithm of \cite{HKPV06}; see \cite{Lavancier, affandi2013approximate}. 
We do not discuss these kinds of approximation here, but as a second contribution, we rather focus
on important classes of DPPs for which the spectral representation can be obtained explicitly on
some subset $S$.
In particular, we investigate in details in Section~\ref{sec:ginibre} the $\beta$-Ginibre process  \cite{goldman2010,deng2014}, a  generalisation of the famous standard Ginibre process \cite{ginibre1965}.
For this model, we provide a simulation algorithm based on the exact spectral representation of the kernel, and we compare it to the simulation based on the eigenvalues of a random matrix (sometimes referred to as the truncated $\beta$-Ginibre approximation). 
Moreover, we also derive in Section~\ref{sec:others} the spectral representation of Gaussian kernels and Bessel-type kernels, two widely used parametric models in spatial statistics, opening the way of perfect simulation for these models.

All algorithms presented in Sections~\ref{sec:projection} and~\ref{sec:ginibre} are available
  in an R-package \citep{Rsoftware} currently available on
  \url{https://github.com/rubak/dppsim} and planned for submission to the Comprehensive R
  Archive Network (CRAN).
The package includes functions for simulation and conditional simulation of any projection DPP, and simulation of the $\beta$-Ginibre process by the spectral algorithm and the truncated $\beta$-Ginibre approximation.

 \section{Preliminaries}\label{preliminaries}
Throughout the manuscript the notation $|.|$ has different meanings: for a complex number $z\in\C$, $|z|$ denotes its modulus; for a vector $x$,  $|x|$ is its infinite norm;  if $J$ is a finite set, then $|J|$ stands for its cardinality; if $S$ is an infinite set, then $|S|$ is its volume.
On the other hand, $\|.\|$ will either refer to the vectorial $\ell^2$  norm, or to the functional $L^2$ norm, depending on the context.
For two vectors $x$ and $y$, $x\cdot y$ denotes their inner product.
For a matrix $A$, we denote by $A'$ its transpose and by $A^*$ its conjugate transpose.

  \subsection{DPPs and basic properties}
 Let $X$ be a point process on $\R^d$ (we refer to \cite{moeller:waagepetersen:04} for background material on spatial point processes). 
 If there exists a non-negative function $\rho_n:(\R^d)^n\rightarrow\R$ such that
$$\E \sum^{\neq}_{x_1,\dots,x_n\in X} f(x_1,\dots,x_n) =\int_{(\R^d)^n}f(x_1,\dots,x_n)\rho_n(x_1,\dots,x_n)\der x_1\cdots\der x_n,$$
for all locally integrable functions $f:(\R^d)^n\rightarrow\R$, where the symbol $\neq$ means that the sum is done for distinct $x_i$, then $\rho_n$ is called the $n$-th order joint intensity function of $X$.
Intuitively, $\rho_n(x_1,\dots,x_n)$ is the infinitesimal probability that $X$ contains  points (among others) located at $x_1,\dots,x_n$.

We say that $X$ is a DPP on $S\subseteq \R^d$  with kernel
$K:S \times S \rightarrow \C$ if for all $n \ge 1$, the joint intensity $\rho_n$ of $X$ exists
and is of the form
\begin{equation}\label{defDPP-S} \rho_n(x_1,\ldots,x_n) = \text{det}[K (x_k,x_l)]_{1\leq k,l\leq n}  \end{equation}
for all $\{x_1,\ldots,x_n \} \subset S$, where
$[K (x_k,x_l)]_{1\leq k,l\leq n}$ denotes the matrix with entries $K(x_k,x_l)$.

Some conditions on $K$ are necessary to ensure the existence of $X$, which depend on whether $X$ is
defined on all of $\R^d$ or a compact subset $S\subset\R^d$.
First, assume that $S$ is a compact set and that $K$ is Hermitian and continuous on $S\times S$, then by
Mercer's Theorem it admits the spectral representation
\begin{equation}\label{decomp1}
K(x,y)=\sum_{k\geq 1} \lambda_k \Phi_k(x)\widebar\Phi_k(y),\quad x,y\in S,
\end{equation}
for some eigenvalues $\lambda_k$ and orthonormal eigenfunctions $\Phi_k$ on $S$, for $k=1,2,\dots$.
(In typical applications $k\in\Z^d$, but for convenience we use a natural number index here.)
A necessary and sufficient condition for existence of a DPP $X$ on $S$ with kernel $K$ is that $\lambda_k\in[0,1]$ for all $k=1,2,\dots$.
We refer the reader to  \cite{HKPV06} for details and more results.
Second, assume that $K$ is invariant under translation on $\R^d$, i.e.\ $K(x,y)=K_0(x-y)$ for some function $K_0$, then a sufficient condition for existence of the DPP with kernel $K$ on $\R^d$ (and then on any subset $S\subset\R^d$) is that $K_0$ is a continuous covariance function on $\R^d$ whose Fourier transform is less than 1; see \cite{Lavancier}.
Note that to check the latter condition it is not necessary to know the spectral representation \eqref{decomp1}
of $K$ on any subset $S$.
In the following, we will assume that the considered DPPs exist (conditions will be recalled for each specific example).

Let us recall some important properties of DPPs \cite{Lavancier}. 
\begin{enumerate}
\item The (first order) intensity of a DPP $X$ on $S$ is by definition $\rho(x)=K(x,x)$, $x\in S$.
    We say that $X$ is homogeneous if $\rho(x)$ is constant. 

\item Its pair correlation function, defined by $g(x,y)=\rho_2(x,y)/(\rho(x)\rho(y))$, reads 
$$g(x,y)=1-\frac{|K(x,y)|^2}{K(x,x)K(y,y)},\quad x,y\in S,$$
whenever $K$ is Hermitian, which we assume henceforth.
The fact that $g(x,y)\leq 1$ reflects that DPPs are models for repulsiveness in a point pattern.

\item Suppose we apply an independent thinning of $X$ with retention probabilities $p(x)$, $x\in S$, then the resulting process is still a DPP with kernel $K(x,y)\sqrt{p(x)p(y)}$.
    In particular, the restriction of $X$ to any subset $S'\subset S$ is the DPP with kernel $K(x,y)\1_{x\in S'}\1_{y\in S'}$.

\item Any smooth transformation of $X$ remains a DPP.
    For example, let $T(x)=Ax+b$, $x\in\R^d$, be an affine transformation on $\R^d$ with $\det(A)\neq 0$,  then $T(X)$ is the DPP on $T(S)$ with kernel $K(A^{-1}(x-b),A^{-1}(y-b))/|\det(A)|$.
\end{enumerate}

As mentioned in the introduction, an important special class of DPPs is the class of projection
DPPs on bounded $S\subset\R^d$.
Their kernel $K$ satisfies $\int_S K(x,z)K(z,y)dz= K(x,y)$ and they generate a fixed number of $n$ points in $S$, where $n=\int_S K(x,x)dx$.
Their eigenvalues in \eqref{decomp1} take only two possible values, 0 or 1,  exactly $n$ of them being 1, the others being 0.  

\medskip

Assume now that all eigenvalues of $K$ in \eqref{decomp1} are strictly less than 1.
Then, the probability density that $X$ has $n$ points located at $x_1,\dots,x_n$ (sometimes called likelihood or Janossy density of $X$) is proportional to $ \text{det}[L (x_k,x_l)]_{1\leq k,l\leq n}$, where $L$ is the kernel defined by $$L(x,y)=\sum_{k\geq 1} \frac{\lambda_k}{1-\lambda_k} \Phi_k(x)\widebar\Phi_k(y),\quad  x,y\in S.$$
 It is common in  the machine learning community to define $X$ from this $L$-kernel, instead of its $K$-kernel, meaning that $X$ is defined through its Janossy densities instead of its joint intensities $\rho_n$.
 If $\lambda_k<1$ for all $k$, both point of views are equivalent and the spectral representation of $K$ can be deduced from that of $L$, and vice-versa. 
On one hand, the advantage of defining $X$ from $L$ is that the eigenvalues of $L$ are not restricted to be less than 1 for $X$ to be well defined and the likelihood function is easier to understand. 
On the other hand, defining $X$ through $L$ implies that the moments of $X$ become unknown, contrary to  (i) and (ii) above. 
Moreover some DPPs, including projection DPPs for which $\lambda_k\in\{0,1\}$, cannot be defined through a $L$-kernel. Note however that in \cite{tremblay2023}, extended $L$-ensembles are introduced and fill this gap (the main setting of that paper concerns discrete DPPs but it is is argued in conclusion that the extension to the continuous setting is straightforward).
In this paper we start from the $K$-kernel and we consider the simulation of $X$ given $K$.

\subsection{Overview of simulation methods}

\subsubsection{Projection DPPs}

\begin{figure}
  \begin{subfigure}[b]{0.5\textwidth}
    \includegraphics[width=\textwidth]{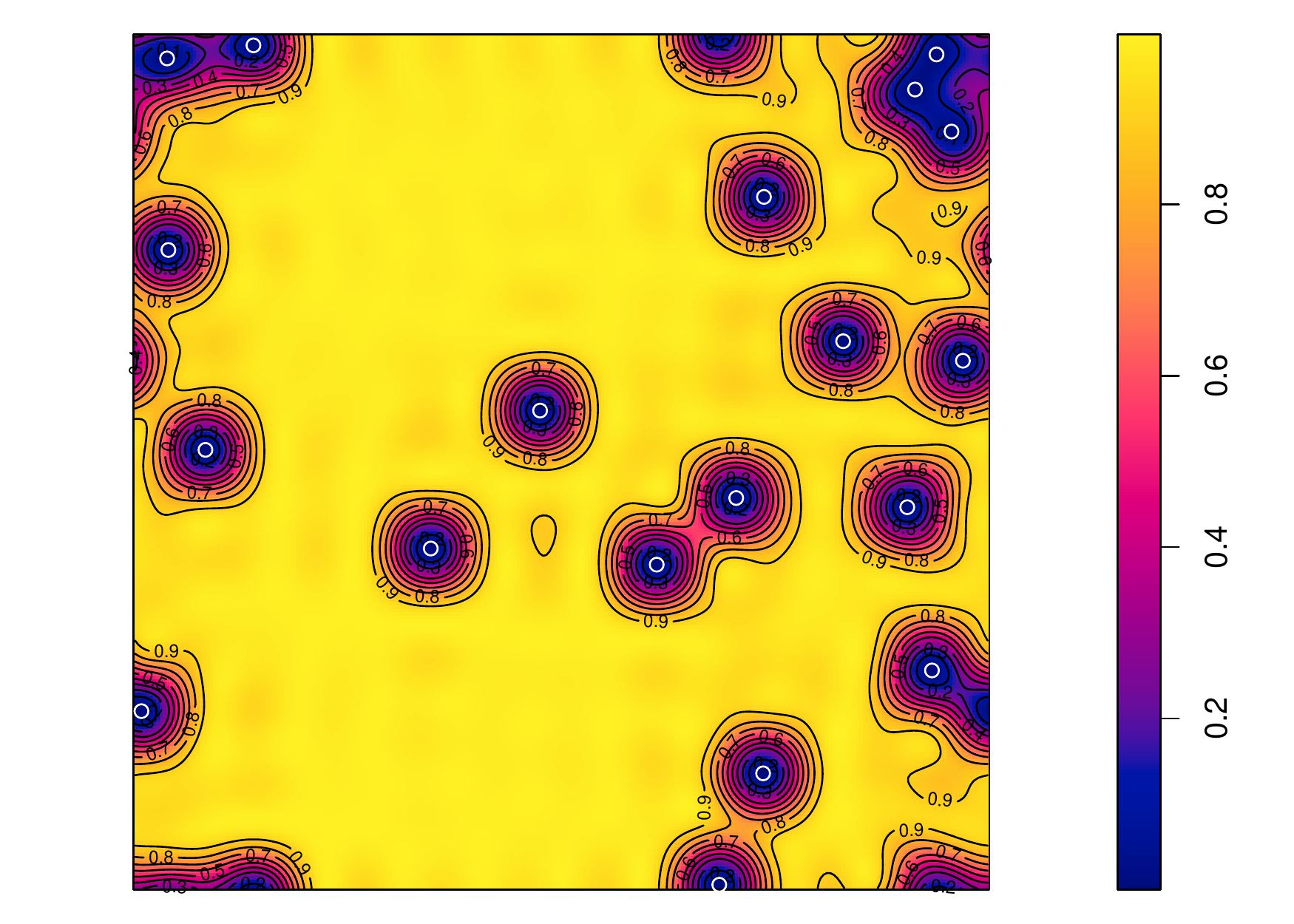}
  \end{subfigure}
  \begin{subfigure}[b]{0.5\textwidth}
    \includegraphics[width=\textwidth]{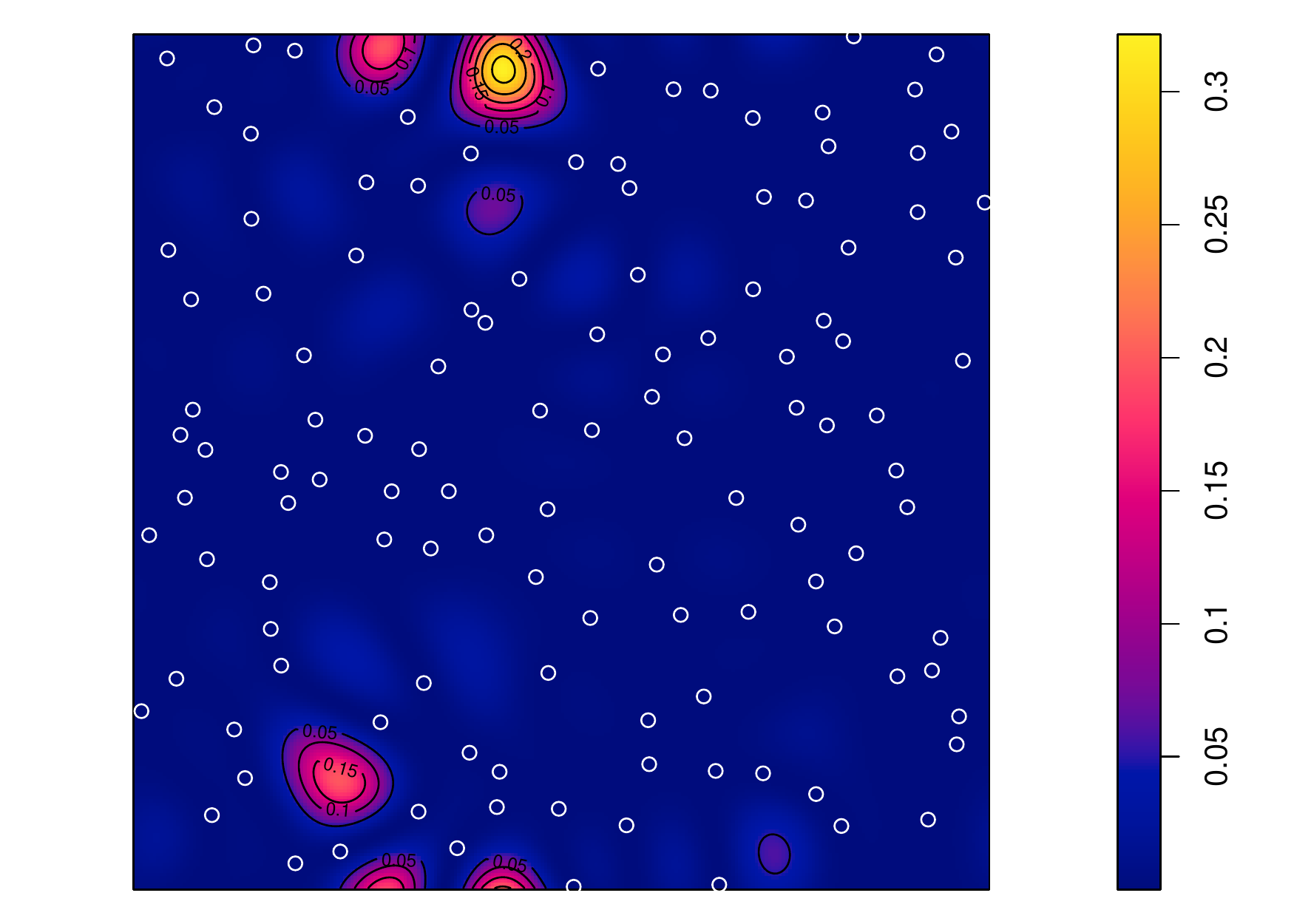}
  \end{subfigure}
  \vskip.5\baselineskip
  \begin{subfigure}[b]{0.5\textwidth}
    \includegraphics[width=\textwidth]{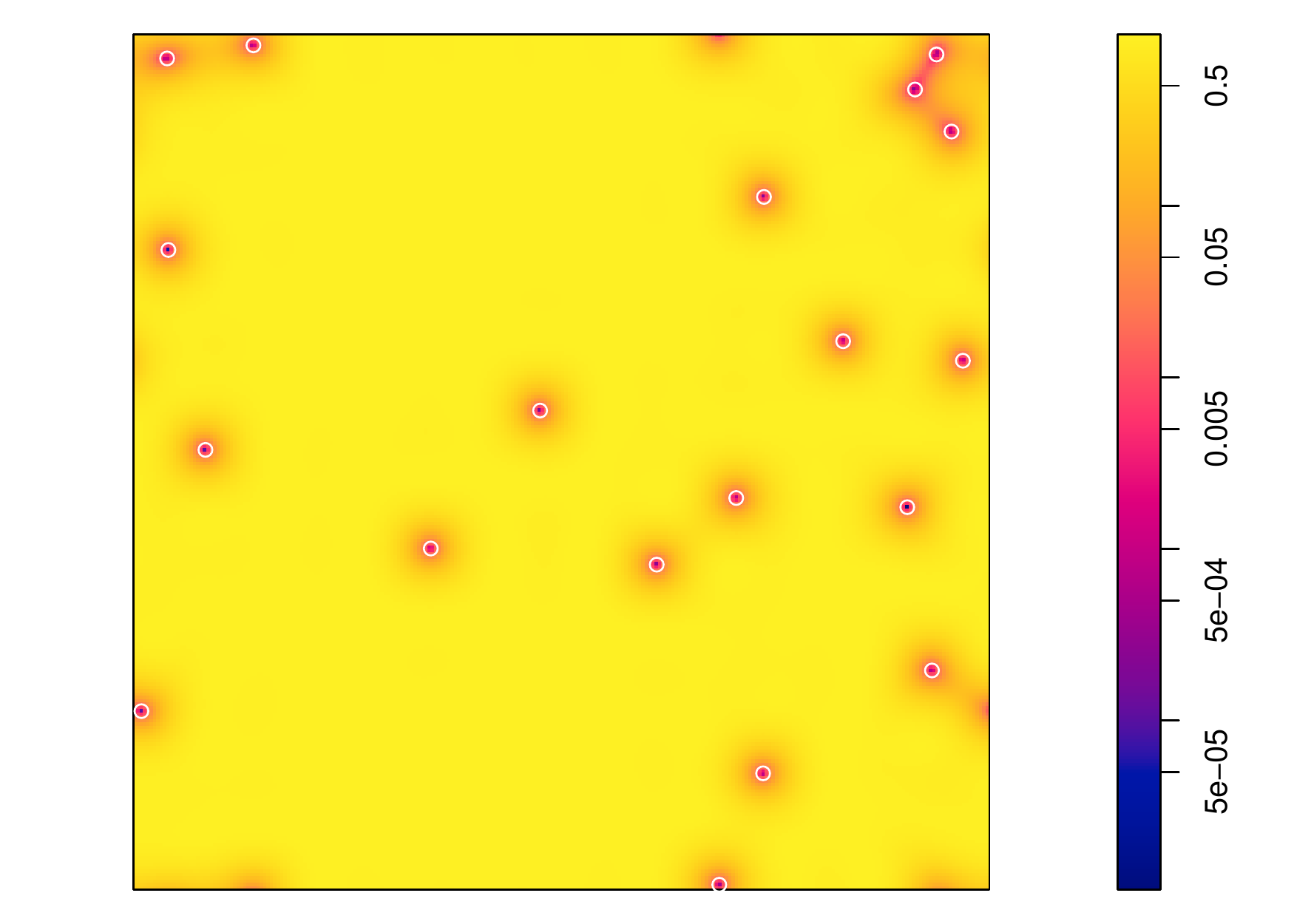}
  \end{subfigure}
  \begin{subfigure}[b]{0.5\textwidth}
    \includegraphics[width=\textwidth]{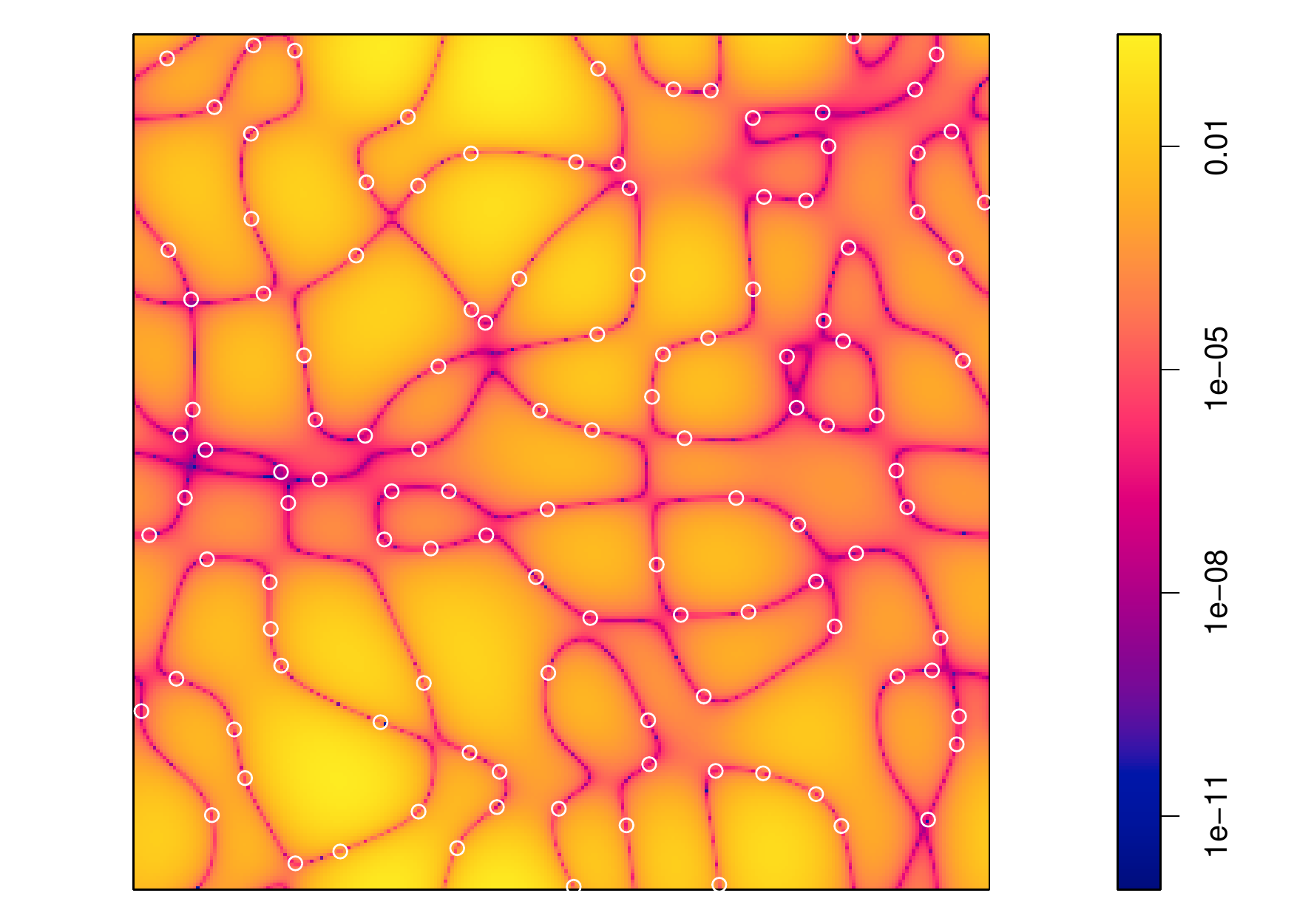}
   \end{subfigure}
\caption{Conditional density $p_i$ at intermediate steps of Algorithms~\ref{alg:nospectral}
  and~\ref{alg:sim}, for the Fourier basis and $n=121$ on the unit square.
  Top row is untransformed values while the bottom row is in log scale.
  Left: density of the 20th point given the 19 first ones (superimposed as white circles);
  right: density of the last point given the preciding 120 points.
}
\label{density and snakes}
\end{figure}

For projection DPPs on a compact set $S\subset\R^d$, perfect simulation algorithms are available, whatever the spectral representation of the kernel is known or not.
We detail these algorithms in Section~\ref{basic algo}.
The idea is to generate the first point of the DPP with respect to the unnormalised density $K(x,x)$ on $S$, then  to generate the second point given the first one with respect to the associated conditional density on $S$, and so on.
The difficulty in this procedure is to be able to simulate, at each step, a point with respect to the conditional density given the previous points. 
Figure~\ref{density and snakes} shows examples of such conditional densities at intermediate steps
of the algorithm when simulating a homogeneous projection DPP having 121 points on the unit square with $K(x,y)=\sum_{|j|\leq 5} e^{i j \cdot (x-y)}$ for $x,y \in [0,1]^2$ and $j\in\Z^2$ (where $i=\sqrt{-1}$ and the unit square in $\R^2$ is identified with the unit square in $\C$).
The top left hand plot is the density of the 20th point given the 19 first points while the top right hand
plot shows the density of the last (121st) point given the first 120 points. The bottom row plots
are the same plots on log scale.
The simulation with respect to these conditional densities is  commonly achieved by rejection
sampling where the proposal is the unnormalised density $K(x,x)$. This procedure can be extremely
costly in the last steps of the algorithm where many rejections may occur due to the complicated shape of the conditional density. 

Currently there seems to be no general way to avoid the costly rejection sampling step, but in
Section~\ref{sec:fourier} we show how to significantly accelerate it in the important particular case where the eigenfunctions of $K$ correspond to the Fourier basis functions.
Moreover, the fact that it is possible to simulate a projection DPP without knowing its spectral representation opens the way to perform conditional simulation, including in-painting,  as we discuss it in Section~\ref{conditional simu}.

\subsubsection{Non-projection DPPs}

If the DPP is not a projection DPP, then the standard algorithm to generate it on $S$, due to \cite{HKPV06}, starts from the spectral representation \eqref{decomp1} of $K$ on $S$.
It consists in first, generating a sequence of Bernoulli variables $B_k$ with respective rates $\lambda_k$ (note that only a finite number of $B_k$'s are non-zero since $\sum_{k\geq 1} \lambda_k<\infty$), and second, given $(B_k)_{k\geq 1}$, generating the projection DPP with kernel  $\sum_{k\geq 1} B_k \Phi_k(x)\widebar\Phi_k(y)$.

In Sections~\ref{sec:ginibre} and \ref{sec:others}, we show that the exact spectral representation
\eqref{decomp1} of the $\beta$-Ginibre kernel, the Gaussian kernel, and the Bessel-type kernel is
available on some sets $S$, opening the possibility of perfect simulation for the associated DPPs
on these sets (and any subset thereof).
We in particular treat in detail the $\beta$-Ginibre process, for which we provide efficient simulation algorithms in Section~\ref{sec:ginibre}.  

However, for most kernels and sets $S$, the expansion \eqref{decomp1}  is not known.
To overcome this issue we may:
\begin{itemize}
\item Consider an approximation of  \eqref{decomp1}. 
For any translation invariant kernel on a rectangular set $S$, a Fourier series approximation is proposed in 
 \cite{Lavancier}, and when the DPP is defined through the $L$-kernel, a Nystr\"om based approximation is suggested in \cite{affandi2013approximate}.

\item Using property (iv) of the previous section, consider a smooth transformation $Y=T(X)$, where the spectral representation of $Y$ on $T(S)$ is known or can be approximated, then generate $Y$ on $T(S)$ and deduce $X=T^{-1}(Y)$.
    In particular, by this property, the simulation of a DPP on any rectangular (resp.\ any
    ellipsoid) set $S$ boils down to the simulation on the unit square (resp.\ on the unit disc).

\item Using property (iii), view $X$ as an independent thinning of some DPP $Y$ whose spectral representation is known or can be approximated, then generate $Y$ and get $X$ by thinning.
    This includes in particular the simulation of $X$ on any subset $S'$ of $S$, provided we know how to generate $X$ on $S$.
    This strategy is also possible for any non-stationary DPP $X$ whose kernel reads $K_0(x-y)\sqrt{\rho(x)\rho(y)}$ with $K_0(0)=1$, provided $\rho(x)$ is uniformly bounded on $S$.
\end{itemize}

If $X$ is defined through its $L$-kernel, the simulation challenge is similar: one needs to know or approximate the spectral representation of $L$ in order to deduce $K$ and use the procedure above.
Let us however mention an alternative solution that does not require the spectral representation of $L$: knowing $L$ on $S$ allows one to have access to the Papangelou intensity of $X$ on $S$.
This opens the possibility to use a CFTP algorithm to simulate perfectly $X$ on $S$, provided $\int_S L(x,x)dx<\infty$, as considered in \cite[Section~5.2]{Decreusefond2016}.
Unfortunately this procedure can  be very slow.

\section{Simulation of a projection DPP}\label{sec:projection}

\subsection{Basic algorithms}\label{basic algo}

Assume that $X\sim DPP(K)$ is a projection DPP with cardinality $n$ on a compact set $S\subset\R^d$.
Then we can simulate  $X$ on $S$ with Algorithm~\ref{alg:nospectral} below. 
This algorithm is justified in \cite{HKPV06}; see also \cite[Section 3.5]{Lyons} and Appendix~\ref{appendix conditional}.
Note that explicit knowledge of the spectral representation \eqref{decomp1} of $K$ is not required.  
This is useful in some situations where we know that $DPP(K)$ exists but no spectral
representation, and even no accurate approximation of it, is available for $K$. This will be the case for the kernel $K$ involved in the inpainting conditional simulation of Section~\ref{conditional simu}.

\begin{algorithm}
\caption{Projection DPP without using the spectral representation}
\label{alg:nospectral}
\begin{algorithmic}
  \STATE {\bf sample} $X_n$ from the distribution with density
  $p_n(x)=K(x,x)/n$, $x\in S$
    \FOR{$i=(n-1)$ to $1$}
      \STATE {\bf set}  ${\bf k}^*_{i}(x)=(K(x,X_n),\dots,K(x,X_{i+1}))$ and ${\bf K}_{i}=(K(X_j,X_l))_{n\geq j,l\geq i+1}$
	\STATE {\bf sample} $X_i$ from the distribution with density
  	\begin{equation}\label{pi nospectral}p_i(x)=\frac 1i\left[K(x,x)-{\bf k}^*_{i}(x) {\bf K}_{i}^{-1} {\bf k}_{i}(x)\right],\quad x\in S \end{equation}
\ENDFOR
  \STATE {\bf return} $\{X_1,\dots,X_n\}$
\end{algorithmic}
\end{algorithm}

On the other hand, if we have access to the spectral representation of $K$,  the conditional density $p_i$ in \eqref{pi nospectral} 
 can be recast to get the more efficient Algorithm~\ref{alg:sim}.
 Some justification for this is given in Appendix~\ref{appendix conditional} and more details can be found in \cite{Lavancier}.
 Here, we assume without loss of generality that the spectral representation of $K$ reads 
\begin{equation}\label{spec_algo}
K(x,y)=\sum_{k=1}^n \Phi_k(x)\bar\Phi_k(y),\quad x,y\in S,
\end{equation}
and we set $\bv(x)=(\Phi_1(x),\dots,\Phi_n(x))'$.

\begin{algorithm}
\caption{Projection DPP using the spectral representation \eqref{spec_algo}}
\label{alg:sim}
\begin{algorithmic}
  \STATE {\bf sample} $X_n$ from the distribution with density
  $p_n(x)=\|\bv(x)\|^2/n$, $x\in S$
  \STATE {\bf set} $\be_1=\bv(X_n)/\|\bv(X_n)\|$
  \FOR{$i=(n-1)$ to $1$}
\STATE {\bf sample} $X_i$ from the distribution with density
   \begin{equation}\label{eq:p_i}p_i(x)=\frac{1}{i}\left[\|\bv(x)\|^2-\sum_{j=1}^{n-i}|\be_j^*\bv(x)|^2\right],\quad x\in S \end{equation}
  \STATE {\bf set} $\bw_i=\bv(X_i)-\sum_{j=1}^{n-i}\left(\be_j^*\bv(X_i)\right)\be_j$, $\be_{n-i+1}=\bw_i/\|\bw_i\|$
  \ENDFOR
  \STATE {\bf return} $\{X_1,\dots,X_n\}$
\end{algorithmic}
\end{algorithm}

To appreciate why   the computation of $p_i$ is faster with Algorithm~\ref{alg:sim} than with Algorithm~\ref{alg:nospectral}, take for instance the evaluation of the second term in \eqref{pi nospectral}.
For each proposed $x$ it requires the computation of ${\bf k}_{i}(x)$ that itself requires $n-i$ evaluations of the vector $\bv(x)$ along with the $n-i$ vectors $\bv(X_j)$ for $i+1\leq j \leq n$.
In contrast, only one evaluation of $\bv(x)$ is required for the same term in Algorithm~\ref{alg:sim} since the $\be_j$'s are stored from the previous step.

\medskip 

From a practical point of view, we need to be able to generate a point with respect to $p_i$, whether we use Algorithm~\ref{alg:nospectral} or \ref{alg:sim}.
This is in fact the bottleneck of these simulation methods.
The standard procedure consists in rejection sampling, where the proposal can be deduced from the bound 
$ip_i(x)\leq K(x,x)$.
Two strategies are then possible:
\begin{itemize}

 \item Strategy 1: If we know how to simulate with respect to the distribution on $S$ with density $K(x,x)/n$, then this distribution can be chosen as the proposal.
     Specifically, if $Z$ is a point generated by this proposal and $U\sim\mathcal U([0,1])$ independent of $Z$, we accept  $Z$ if $ip_i(Z)/K(x,x) > U$.
     The probability of acceptance is $i/n$.

\item Strategy 2:  If we only know an upper-bound for $K(x,x)$, i.e. $K(x,x)\leq M$ for all $x\in S$, then $p_i(x)\leq M/i$ and rejection sampling can be applied where the proposal is the uniform distribution on $S$. 
 Specifically, if $Z$ is generated uniformly on $S$ and $U\sim\mathcal U([0,1])$ independent of $Z$, we accept  $Z$ if $ip_i(Z)/M > U$.
 The probability of acceptance is $i/(M|S|)$.
 
  \end{itemize}
The second strategy is always an option in practice, provided we initially approximate the upper-bound $M$ numerically.
Note that given $S$ and $K$, this bound can be found offline and once for all.

\begin{rem}
A natural and commonly used idea in spatial statistics consists in leveraging the local contribution of a point process to generate a new point given the other ones. 
Unfortunately, the conditional density $p_i$ is not well localised, especially in presence of many conditional points, which corresponds to the smallest values of $i$ or equivalently to the last (most time consuming) steps of the simulation algorithms, ruling out the idea of localisation.
This is clearly illustrated in Figure~\ref{density and snakes}: While in the first steps of the
algorithm (left hand plots) $p_i$ seems to be localised around the conditional points, the picture looks different in the right hand plots, where some ``empty'' regions are associated to very small values of the density $p_i$, whereas other seemingly similar regions are associated with a high density.
This effect is particularly evident in the bottom right hand plot on log scale, that shows quite stunning long range dependencies in $p_i$. 
This demonstrates that not only the neighbour points of these regions impact the values of $p_i$, but really the whole point pattern.

\end{rem}

\subsection{Refinement for the Fourier basis}\label{sec:fourier}

The Fourier basis plays an important special role in continuous DPPs for its simplicity, the homogeneity and the repulsiveness it yields \cite{coeurjolly2020monte}, and the fact that it is at the heart of the spectral approximation of invariant-translation kernels in \cite{Lavancier}.
Let us assume for simplicity that $S=[0,1]^d$ (remember that the simulation on any rectangular windows boils down to this case).  
The kernel of a projection DPP on $S$ based on the Fourier basis reads
\begin{equation*}
K(x,y)=\sum_{j\in J}
  \mathrm{e}^{2\mathrm{i}\pi j\cdot (x-y)},
\end{equation*}
where $J\subset \Z^d$ is some finite subset of $\Z^d$ with cardinality $|J|=n$.

 In this case, the more efficient  Algorithm~\ref{alg:sim} is feasible and the two aforementioned strategies for the simulation with respect to $p_i$ are similar since $K(x,x)=n$ (then $M=n$ in Strategy 2).
 The Algorithm builds on rejection sampling with the uniform distribution as a proposal.
 The costly part in this procedure is the evaluation of $p_i(Z)$ for any proposed point $Z$ generated from the uniform distribution on $S$, which becomes all the more problematic in the last steps of the $i$-loop in Algorithm~\ref{alg:sim} when many rejections are expected, the probability of acceptance then being of order $1/n$.  
 
 To accelerate the procedure, we suggest to exploit the following bound; see Appendix~\ref{sec:bound fourier}:
\begin{equation}\label{bound Fourier}
i p_i(x)/n \leq  
     \min_{i+1\leq k\leq n} \left(1, P(x-X_k)/n \right),
\end{equation}
where $$P(x)=4  \pi^2 \sum_{j\in J} (j\cdot x)^{2} - \frac{4\pi^2} n  \left(\sum_{j\in J} j\cdot x \right)^{2}$$ is a second-order polynomial depending on $d$-variables.
Note that the coefficients of $P$ only depend on the index set $J$ and  can thus be computed offline before  running Algorithm~\ref{alg:sim}. 
The inequality \eqref{bound Fourier} provides a simple first bound to test against in the rejecting sampling procedure, avoiding the more costly evaluation of $p_i$ for many proposed points.
The refined rejection sampling procedure is detailed in Algorithm~\ref{alg:rejection}. 


\begin{algorithm}
\caption{Refined rejection sampling for $p_i$ with the Fourier basis}
\label{alg:rejection}
\begin{algorithmic}
\REPEAT
   \STATE {\bf sample} $Z$ from the uniform distribution on $S$
  \STATE {\bf generate} $U\sim\mathcal U([0,1])$ independent of $Z$
  \IF{$\min_{i+1\leq k\leq n} P(Z-X_k)/n < U$}
  \STATE reject $Z$
  \ELSIF{$ip_i(Z)/n > U$}
  \STATE accept $Z$ 
  \ELSE 
  \STATE reject $Z$
    \ENDIF
    \UNTIL $Z$ is accepted 
  \end{algorithmic}
\end{algorithm}

We tested the performance of this refinement by simulation.
Table~\ref{table fourier} reports the fraction of time needed to generate different point patterns
in dimension $d=2$ by using Algorithm~\ref{alg:rejection} compared to using simple rejection
sampling.
The entries are averaged over 500 replications.
The tested models are first, the projection DPP with kernel $K(x,y)=\sum_{|j|\leq \ell}  e^{i j \cdot (x-y)}$,  
that leads to a cardinality of $n=(2\ell+1)^2$ points, which we call ``most repulsive'' kernel with
intensity $\rho=n$ in Table~\ref{table fourier}, and second, the Gaussian-type DPP (see
Section~\ref{sec:gaussian}) with a comparable intensity $\rho=n$ and an interaction parameter
$\alpha=\alpha_{\max}$ (strong repulsion) and $\alpha=\alpha_{\max}/2$ (mild repulsion).
Here $\alpha_{\max}=1/\sqrt{\rho\pi}$ is the maximal possible value of $\alpha$ for a Gaussian-type DPP with intensity  $\rho$, and the simulation exploits the Fourier approximation of \eqref{decomp1} as introduced in \cite{Lavancier}, making the use of Algorithm~\ref{alg:rejection} relevant. 
Table~\ref{table fourier} also reports the ``bound rate'' which summarises how often the bound  \eqref{bound Fourier} was used on
average.
This rate was empirically quite stable across the different values of $\rho$ that we implemented and this is why only a single number is given.
Specifically, a bound rate of 0.40 means
that $40\%$ of the rejections in the algorithm was done directly thanks to the
bound  \eqref{bound Fourier}. 
This rate can be viewed as the maximal theoretical gain we can expect thanks to Algorithm~\ref{alg:rejection}, if the only computation time was due to this rejection step in Algorithm~\ref{alg:sim}.

\begin{table}[ht]
\centering
\begin{tabular}{l|r|rrrrr}
  \hline
&& \multicolumn{5}{c}{Intensity ($\rho$)} \\  
    DPP model & Bound rate & $25$ & 81 & 289 & 625 & 1089 \\ 
  \hline
Most repulsive & 0.41 & 1.10 & 1.00 & 0.96 & 0.77 & 0.72 \\ 
  Gauss ($\alpha = \alpha_{\max}$) & 0.24 & 0.99 & 0.93 & 0.93 & 0.85 & 0.85 \\ 
  Gauss ($\alpha = \alpha_{\max}/2$) & 0.06 & 1.02 & 0.97 & 0.99 & 0.99 & 0.96 \\ 
   \hline
  \end{tabular}
  \caption{For three DPP models (detailed in the text), generated 500 times on $S=[0,1]^2$:
    fraction of rejections due to \eqref{bound Fourier} when using Algorithm~\ref{alg:rejection} (Bound rate) and fraction of time needed for the simulation using the refinement of Algorithm~\ref{alg:rejection} compared with not using it, for different values of the intensity $\rho$ of the DPP models.}
\label{table fourier}  
\end{table}

From Table~\ref{table fourier}, we conclude that using Algorithm~\ref{alg:rejection} in dimension $d=2$ may lead to a speed improvement up to almost $30\%$ (for the most repulsive kernel and about 1000 points), and never really slows down the simulation even when no clear improvement can be expected (as for the mild repulsive Gaussian-type DPP for which the bound rate is only 0.06).
We have carried out similar simulations in dimension $d=1$, not detailed here, where we observed a bound rate of $65\%$ and a gain up to $45\%$ for the most repulsive kernel with 1000 points.
In dimension $d=3$, comparable simulations showed a bound rate of $25\%$ and a speed improvement up to $16\%$ for the most repulsive kernel with 1000 points.
We expect the improvements to be smaller in higher dimensions, making the use of Algorithm~\ref{alg:rejection} superfluous in this case. 

\medskip

Finally, let us mention an additional possible refinement in dimension $d=1$.
In this case, the bound \eqref{bound Fourier} can be manipulated to provide a more efficient  proposal than the uniform distribution, that can be easily simulated by the inversion method.
The details are given in Appendix~\ref{sec:fourierd1}. 
With this new proposal, the acceptance probability in the rejection sampling for the last step of
Algorithm~\ref{alg:sim}, i.e.\ for the simulation of the last point given the $n-1$ first ones, can be $3$ times higher that one associated to a uniform proposal; see Appendix~\ref{sec:fourierd1}. 
Unfortunately, this procedure does not generalise easily to higher dimensions, in the sense that it seems difficult to derive a similar efficient proposal  from \eqref{bound Fourier} in dimension $d>1$.

\subsection{Conditional simulation}\label{conditional simu}

Let again $X$ be a projection DPP on $S\subset\R^d$ with cardinality $n$. 

By construction, the last $i$ steps of Algorithms~\ref{alg:nospectral} and \ref{alg:sim} provide an exact simulation of $X\!\smallsetminus\!\{X_n,\dots,X_{i+1}\}$ given that $\{X_n,\dots,X_{i+1}\}\subset X$.
This conditional simulation may be useful in practice if we observe a point pattern in $S$ with missing points and we wish to generate some realisations of the full point pattern in $S$.
This is possible if we agree that the complete unobserved point pattern can be modelled by a
projection DPP, to be chosen, and that the missing points are due to an independent thinning with known retention probability $q(x)$, $x\in S$.

\begin{figure}
\centering
\includegraphics[scale=0.3]{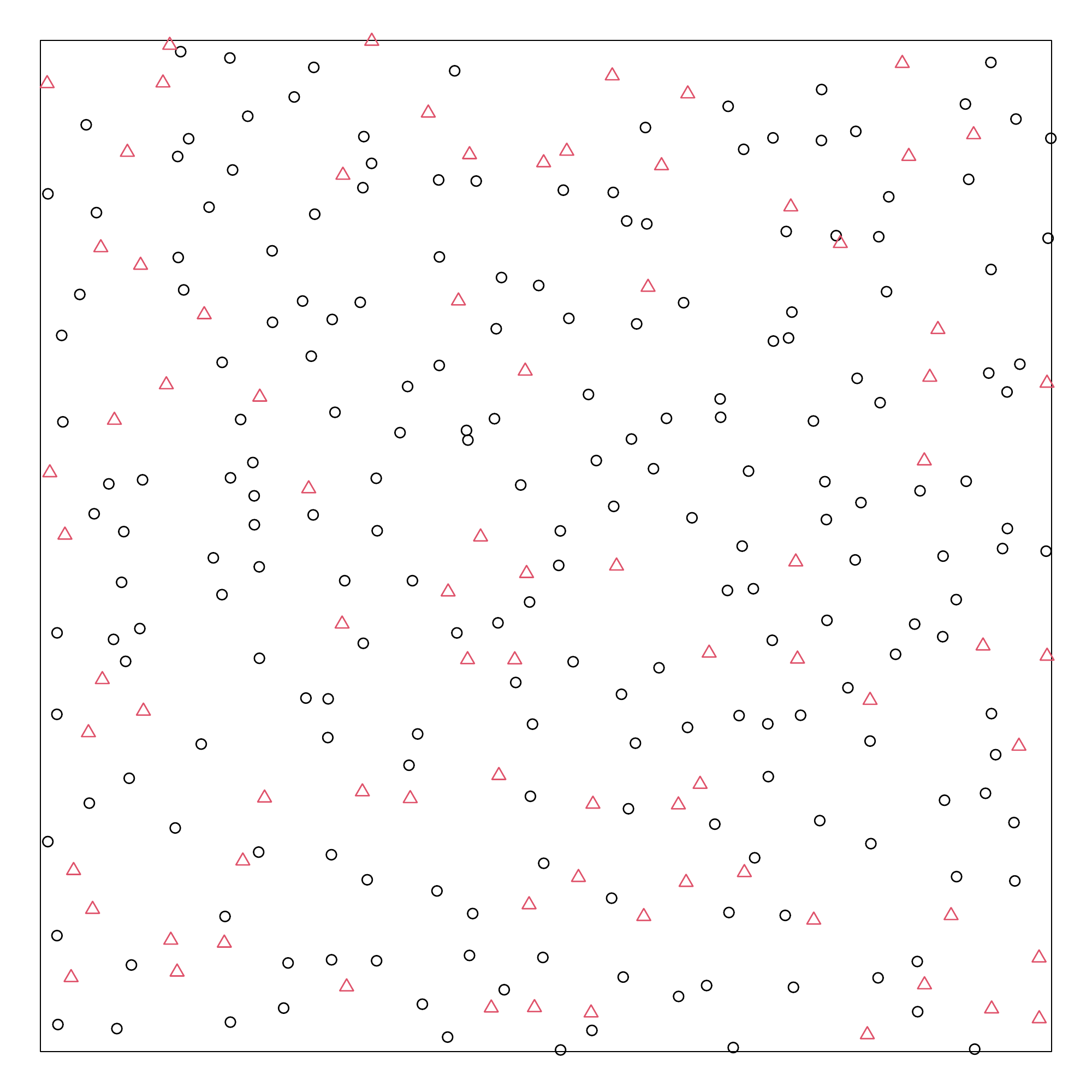}
\includegraphics[scale=0.3]{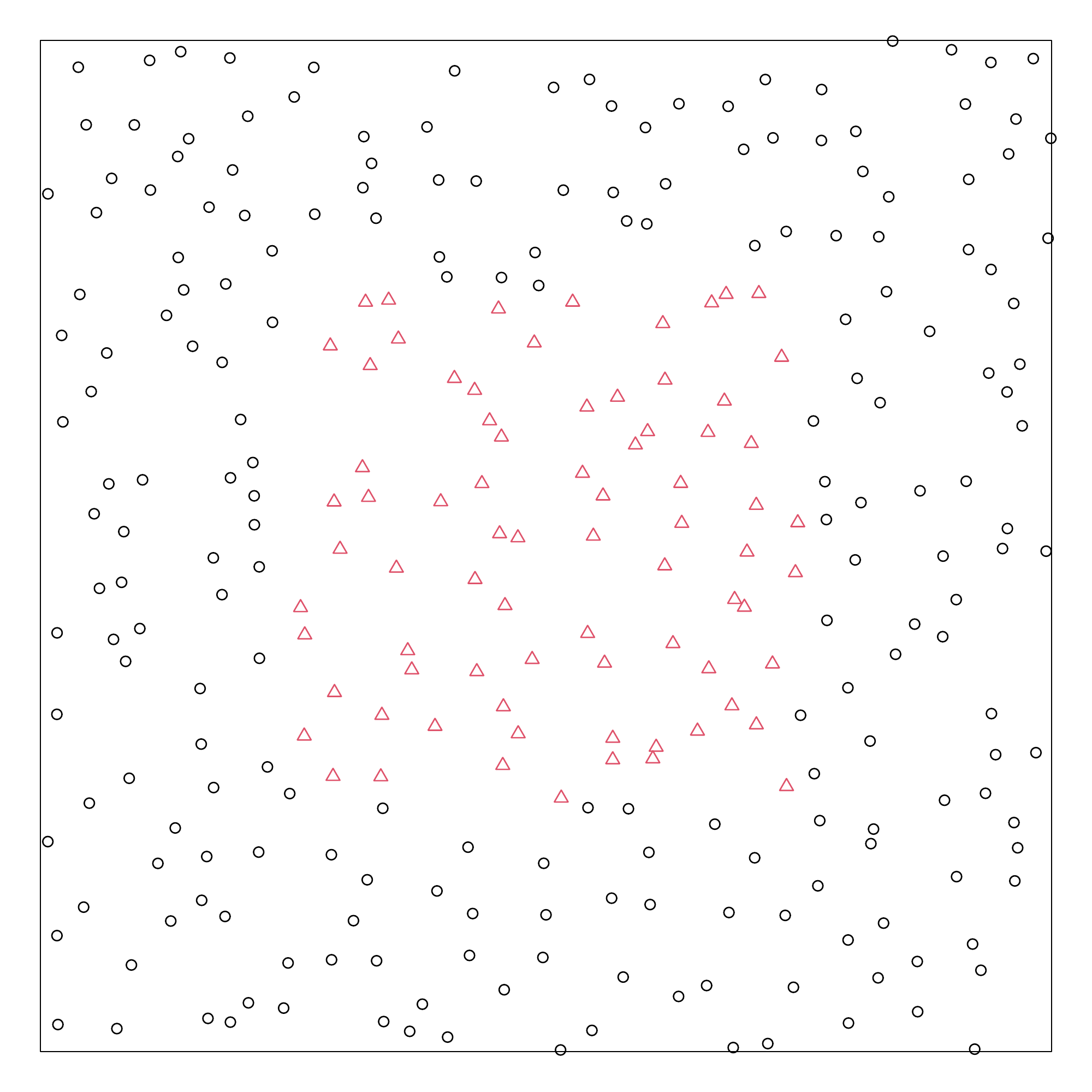}
\caption{Conditional simulation of the red points (shown as triangles) given the black points, assuming that the full point pattern formed by the union of the black and red points is a projection DPP.
Left:  simulation in the whole domain.
Right: simulation in the inner square (in-painting).}
\label{fig:conditional}
\end{figure}

As an example, assume that $X$ is a projection DPP on $S=[0,1]^2$ with kernel $K(x,y)=\sum_{|j|\leq
  \ell}  e^{i j \cdot (x-y)}$, so cardinality $n=(2\ell+1)^2$, where we need to choose $\ell$ (or
equivalently $n$), and we observe only $m$ points of $X$, $m\leq n$, coming from an independent thinning of $X$ with known retention probability $q(x)$, $x\in S$.
Using the relation $\E(m)=\int_S q(x) K(x,x) dx = n\int_S q(x) dx$, we 
deduce a plausible value of $n$ by replacing  $\E(m)$ by the
observed value of $m$. 
Then we can use Algorithm~\ref{alg:sim} to generate the $n-m$ remaining points given the $m$ observed ones. 
This example is illustrated  in the leftmost plot of Figure~\ref{fig:conditional} where the red triangles represent a possible realisation of the unobserved points in $S=[0,1]^2$ given the observed point pattern formed by the black circles.
Here we assumed that $q(x)=1/2$, so that $n$ is twice  the number of observed black points.

We chose in the above example a specific form of $K$, up to $n$, that may not be relevant for other applications.
A general procedure to select a projection DPP $X$ that fits well with the observed conditional points is discussed in Appendix~\ref{choice proj dpp}.

\medskip

Alternatively, we may consider in-painting conditional simulation, that is the simulation of $X\cap A$ given that $X\cap A^c=\{X_1,\dots,X_{m}\}$, $m\leq n$, where $A$ and $A^c$ constitute a partition of $S$.
(Here we agree that if $m=0$, $X\cap A^c=\emptyset$.)
Since $X$ is a projection DPP, it is easily verified (see Appendix~\ref{outside set}) that the latter conditional point process is a projection DPP with cardinality $n-m$ and  kernel, when $m>0$:  
\begin{equation}
K_{X_1,\dots,X_{m}}(x,y)=K(x,y)-{\bf k}^*_{m}(x) {\bf K}_{m}^{-1} {\bf k}_{m}(y),\quad x,y\in A,\label{eq:cond_kernel}
\end{equation}
where ${\bf k}^*_{m}(x)=(K(x,X_1),\dots,K(x,X_{m}))$ and ${\bf K}_{m}=(K(X_j,X_l))_{1\leq j,l\leq m}$.  
If $m=0$, $K_\emptyset(x,y)=K(x,y)\1_A(x)\1_A(y)$.
The spectral representation of the conditional kernel \eqref{eq:cond_kernel} is unlikely to be known, even if the spectral representation of $K$ on $S$ is explicitly given.
Nonetheless the simulation can be done by Algorithm~\ref{alg:nospectral}.

As an illustration,  we have generated  in the rightmost plot of Figure~\ref{fig:conditional} the red triangles in the inner square  $[1/4,3/4]^2$ given the point pattern in black circles outside this square.
The full point pattern was assumed to be the same DPP as before, where  $n$ was fixed to the number of observed black circles times $4/3$ (that is the ratio of the volume of the full region to the volume of the observed region).
A more adaptive choice of $K$ based on the observed points is discussed in Appendix~\ref{choice proj dpp}.

\section{$\beta$-Ginibre process}\label{sec:ginibre}

The $\beta$-Ginibre process, for $\beta>0$,  is the DPP in the complex plane $\C$ with kernel 
\begin{equation}\label{Ginibre kernel}
K(x,y)=\rho \exp\left(\frac{x\bar y}{\beta}-\frac{|x|^2+|y|^2}{2\beta}\right),\quad x,y\in\C.
\end{equation}
Note that we could equivalently consider this process in $\R^2$ but it is commonly defined in $\C$, a point of view which we will respect in the following.
The $\beta$-Ginibre process is a stationary and isotropic point process with intensity $\rho>0$, that exists if and only if $\rho\beta\pi\leq 1$.
The standard Ginibre process introduced in \cite{ginibre1965} corresponds to the particular case $\rho=1/\pi$ and $\beta=1$.
The $\beta$-Ginibre process, initially considered in \cite{goldman2010} and \cite{deng2014}, can be viewed as an independent  thinning of the Ginibre process with retention probability $\rho\beta\pi$, followed by the homothety with ratio $\sqrt\beta$.
Given $\rho>0$, the $\beta$-Ginibre process includes the Poisson point process with intensity $\rho$ as a limiting case when $\beta\to 0$, and  when $\beta=1/(\rho\pi)$, it is one of the most repulsive stationary DPPs with intensity $\rho$  in the sense of \cite{reilly}.

We discuss and compare in this section two strategies to generate the $\beta$-Ginibre process on the ball $\B_R=\{|x|<R\}$ in $\C$, for any $R>0$.
Remember that the simulation on any compact set $S\subset \B_R$ can then be performed by removing the points in $\B_R\!\smallsetminus\! S$.
The first strategy is a standard procedure that takes advantage of the representation of the Ginibre process as the distribution of certain eigenvalues of an infinite matrix.
The second strategy exploits the (unusual) fact that the spectral decomposition \eqref{decomp1} of $K$ is explicitly known on $\B_R$, and so the spectral algorithm of \cite{HKPV06} can be employed.

\begin{rem}
The second order properties of the Ginibre process, i.e.\ the intensity and the pair correlation function, are similar to those of the Gaussian-type DPP with parameters $\rho$ and $\alpha$ (see Section~\ref{sec:gaussian}) with the correspondence $\beta=\alpha^2/2$.
Note however that the Ginibre process can reach more repulsiveness in view of the existence condition $\beta\leq 1/(\rho\pi)$, while $\alpha^2/2\leq 1/(2\rho\pi)$ is required for existence of the Gaussian-type DPP. 
\end{rem}

\subsection{Approximation by the truncated $\beta$-Ginibre process}\label{sec:ginibre_eigen_alg}

The Ginibre process can be viewed as the distribution of the eigenvalues of an infinite matrix with i.i.d.\ standard complex Gaussian entries \citep{ginibre1965}. 
To simulate (approximately) the Ginibre process on $\B_R$, the standard procedure consists of
generating a finite $n\times n$ matrix of i.i.d. standard complex Gaussians, where $n$ has to be chosen according to $R$, and then computing its $n$ complex eigenvalues.
The simulation of the $\beta$-Ginibre process is easily deduced by the thinning and rescaling steps explained in the beginning of this section.
Given the importance of this procedure, we provide some details below and we discuss the choice of $n$.

The distribution of the $n$ eigenvalues above (followed by the thinning and scaling steps)
corresponds to the distribution of the truncated $\beta$-Ginibre process, which is the DPP on $\C$ whose kernel is the truncation at $k=n-1$ of \eqref{GinibreC} defined in Section~\ref{sec:ginibre_sequential_alg} below, that is
\begin{equation}\label{Ginibre_truncated}
K_n(x,y)=\sum_{k=0}^{n-1} \rho\beta\pi \Phi_k(x)\widebar\Phi_k(y),\quad x,y\in\C.\end{equation}
We refer to \cite{decreusefond_flint_vergne_2015} for some discussion of this process and on other modified versions of the Ginibre process. 
As already noticed in  \cite{ginibre1965}, this DPP is not stationary and is mostly concentrated on the disc with radius $\sqrt{n\beta}$.
Inside this disc the truncated $\beta$-Ginibre process and the genuine $\beta$-Ginibre process are very similar, their differences arising mostly close to the border.
In particular the intensity of the truncated $\beta$-Ginibre process reads
$$\rho(x)=K_n(x,x)=\rho e^{-|x|^2/\beta} \sum_{k=0}^{n-1} \frac{(|x|^2/\beta)^k}{k!},\quad x\in \C.$$
This intensity is always less than $\rho$ (the intensity of the $\beta$-Ginibre process) and we have  for any $ |x|\leq \sqrt{n\beta}$ (see \cite{ginibre1965}):
\begin{equation}\label{error rho}
0\leq\frac{\rho - \rho(x)}{\rho} \leq e^{-|x|^2/\beta} \frac{(|x|^2/\beta)^{n}}{n!} \frac{n+1}{n+1 - |x|^2/\beta} :=M_\beta(|x|,n).
\end{equation}
To simulate approximately the $\beta$-Ginibre process on $\B_R$, the standard procedure consists in simulating the truncated $\beta$-Ginibre process for $n$ large enough so that $R\leq \sqrt{n\beta}$ and so that the approximation is satisfying.
To our knowledge, there is no clear recommendation for the choice of $n$.
Choosing $n=\lceil R^2/\beta\rceil$ would lead to a poor approximation close to the border of the disc.
We instead  suggest to choose $n$ so that the upper bound in \eqref{error rho} is uniformly less than a prescribed error $\epsilon$ on $\B_R$.
Specifically, 
we choose $n=\lceil n_\epsilon\rceil $   
where $n_\epsilon$ is such that $M_\beta(R,n_\epsilon )=\epsilon$.
In the simulations presented in Section~\ref{simu ginibre}, we fixed $\epsilon=10^{-10}$.
The full procedure is summarised in Algorithm~\ref{alg:ginibre_eigen}.

\begin{algorithm}
\caption{$\beta$-Ginibre process on $\B_R$ by the eigenvalues method}
\label{alg:ginibre_eigen}
\begin{algorithmic}
  \STATE {\bf set} $n=\lceil n_\epsilon\rceil$ where $n_\epsilon$ solves $M_\beta(R,n_\epsilon )=\epsilon$; see \eqref{error rho}  
  \STATE {\bf sample} the matrix $M=[(A_{k,l} +\dot\iota B_{k,l})\sqrt{\beta/2}]_{1\leq k,l\leq n}$ where $\dot\iota=\sqrt{-1}$ and $A_{k,l}$ and $B_{k,l}$ are all independent $\mathcal N(0,1)$ random variables
  \STATE {\bf compute} the eigenvalues $\{Z_1,\dots,Z_n\}$ of $M$
   \STATE {\bf set} $X=\{\emptyset\}$
    \FOR{$i=1$ to $n$}
    \STATE  {\bf generate} $U\sim\mathcal U([0,1])$ 
    \IF{$U<\rho\beta\pi$ and $|Z_i|<R$}
    \STATE $X \leftarrow X\cup \{Z_i\}$
    \ENDIF
\ENDFOR
\STATE {\bf return} $X$
  \end{algorithmic}
\end{algorithm}

\subsection{Spectral algorithm}\label{sec:ginibre_sequential_alg}

Starting from $\sum_{k\geq 0} (x\bar y /\beta)^k/k! = \exp(x\bar y/\beta)$, it is not difficult to verify that the spectral representation of the $\beta$-Ginibre kernel \eqref{Ginibre kernel} on $\C$ reads
\begin{equation}\label{GinibreC}
K(x,y)=\sum_{k\geq 0} \rho\beta\pi \Phi_k(x)\widebar\Phi_k(y),\quad x,y\in\C,\end{equation}
where $$\Phi_k(x)=\frac{x^k}{\sqrt{\pi k!\beta^{k+1}}}e^{-|x|^2/(2\beta)},\quad x\in\C.$$
As noticed in \cite{decreusefond_flint_vergne_2015}, the functions $\Phi_k$ satisfy the unusual property to remain orthogonal on the ball $\B_R$ for any $R>0$.
So the spectral representation of $K$ on $\B_R$ is simply the restriction of \eqref{GinibreC} to $\B_R$:
$$K(x,y)=\sum_{k\geq 0} \rho\beta\pi \|\Phi_k\|^2_{R}\frac{\Phi_k(x)}{\|\Phi_k\|_{R}}\frac{\widebar\Phi_k(y)}{\|\Phi_k\|_{R}},\quad x,y\in\B_R,$$
where $ \|\Phi_k\|^2_{R}=\int_{\B_R} |\Phi_k(x)|^2dx$ is introduced to normalise the eigenfunctions on $\B_R$.
Some algebra yields $\|\Phi_k\|^2_{R}=\gamma(k+1,R^2/\beta)/k!$ where $\gamma(a,z)=\int_0^z t^{a-1}e^{-t}dt$ is the incomplete gamma function.
Finally the spectral representation of $K$ on $\B_R$ is
\begin{equation}\label{spectral Ginibre}
K(x,y)=\sum_{k\geq 0} \lambda_k \tilde\Phi_k(x)\widebar{\tilde\Phi}_k(y),\quad x,y\in\B_R,
\end{equation}
where \begin{equation}\label{lambda Ginibre}\lambda_k=\rho\beta\pi\gamma(k+1,R^2/\beta)/k!\end{equation} and 
 \begin{equation}\label{phi Ginibre}\tilde\Phi_k(x) = \frac{x^k}{\sqrt{\pi\beta^{k+1}\gamma(k+1,R^2/\beta)}}e^{-|x|^2/(2\beta)},\quad x\in\B_R.\end{equation}

 To apply the spectral algorithm of \cite{HKPV06}, we need to first simulate a sequence of Bernoulli random variables $B_k$, with respective parameters $\lambda_k$, $k\geq 0$.
We know that $M=\max_{k\geq 0} \{B_k\neq 0\}$ is finite because $\sum \lambda_k<\infty$, but the simulation of $M$ is not straightforward; see \cite[Appendix D]{lavancier_extended}.
A simple solution consists in setting a maximal number $k_{\max}$ of Bernoulli variables that we generate.
This amounts to truncate the representation \eqref{spectral Ginibre} to $k_{\max}-1$, which gives exactly  the same kernel as in \eqref{Ginibre_truncated} but restricted to $\B_R$.
A clever choice of $k_{\max}$ can then be carried out as for the choice of $n$ in  \eqref{Ginibre_truncated}, that is $k_{\max}=\lceil k_\epsilon\rceil $   
where $k_\epsilon$ is the solution of $M_\beta(R,k_\epsilon )=\epsilon$.
Here $\epsilon$ is a prescribed error ($\epsilon=10^{-10}$ in Section~\ref{simu ginibre}) that guarantees through \eqref{error rho} that the choice of $k_{\max}$ has a little impact on the expected number of points. 
The details of this spectral approach are given Algorithm~\ref{alg:ginibre_spectral}.

\begin{algorithm}
\caption{$\beta$-Ginibre process on $\B_R$ by the spectral method}
\label{alg:ginibre_spectral}
\begin{algorithmic}
  \STATE {\bf set} $k_{\max}=\lceil k_\epsilon\rceil$ where $k_\epsilon$ solves $M_\beta(R,k_\epsilon )=\epsilon$; see \eqref{error rho}  
  \STATE {\bf sample} independently $B_0\sim \mathcal B(\lambda_0),\dots,B_{k_{\max}-1}\sim \mathcal B(\lambda_{k_{\max}-1})$ where the $\lambda_k$'s are given by \eqref{lambda Ginibre}
   \IF{$B_k=0$ for all $k=0,\dots,k_{\max}-1$}
  \STATE {\bf return} the empty point configuration $\{\emptyset\}$
  \ELSE 
  \STATE Given $B_0,\dots, B_{k_{\max}-1}$, generate the projection DPP with kernel 
  $$\sum_{k=0}^{k_{\max}-1} B_k \tilde\Phi_k(x)\widebar{\tilde\Phi}_k(y),\quad x,y\in\B_R,$$
  using Algorithm~\ref{alg:sim}, where the $\tilde\Phi_k$'s  are given by \eqref{phi Ginibre}

    \ENDIF
   \end{algorithmic}
\end{algorithm}

\subsection{Simulation study}\label{simu ginibre}

\begin{figure}
  \begin{subfigure}[b]{0.5\textwidth}
    \includegraphics[width=\textwidth]{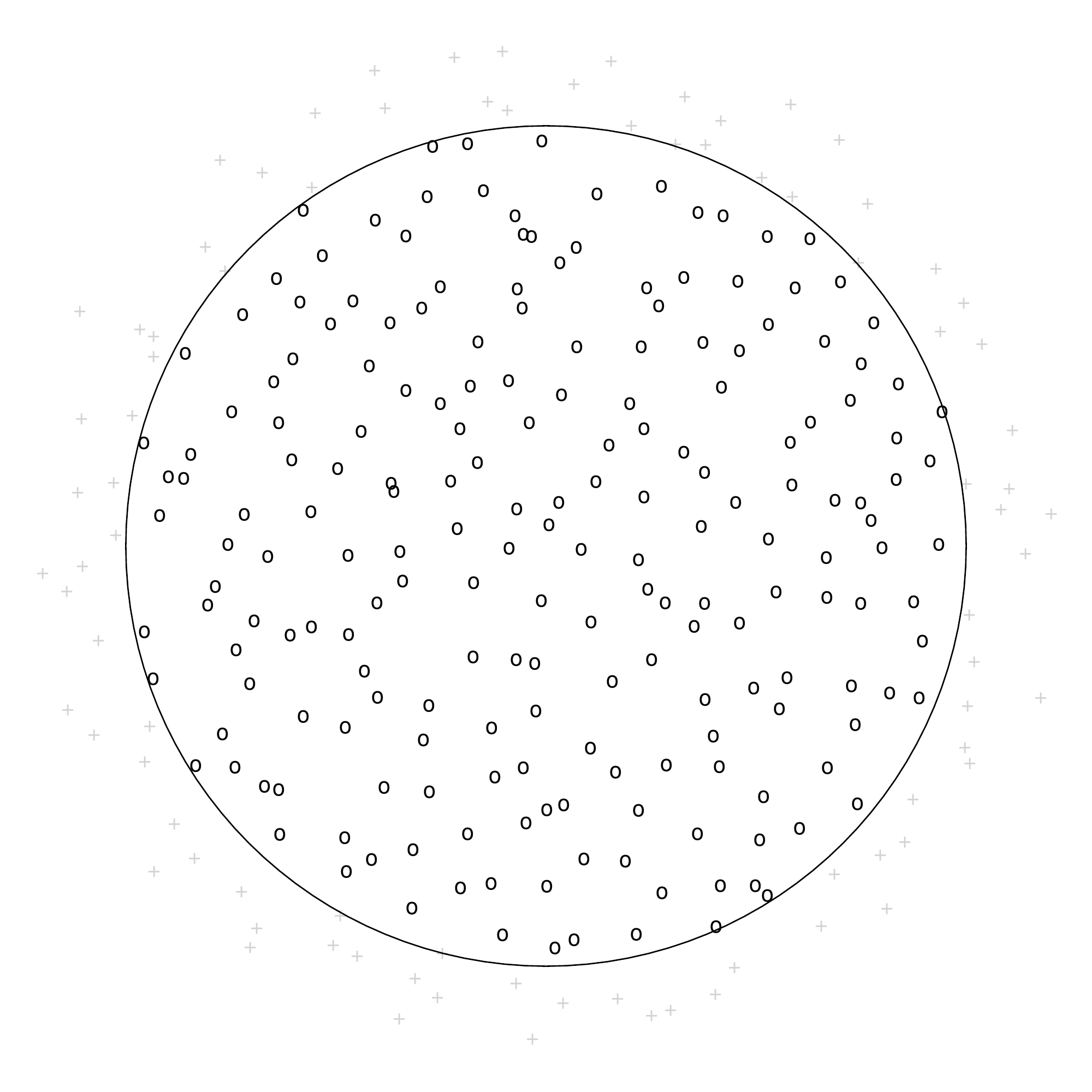}
  \end{subfigure}
  \begin{subfigure}[b]{0.5\textwidth}
    \includegraphics[width=\textwidth]{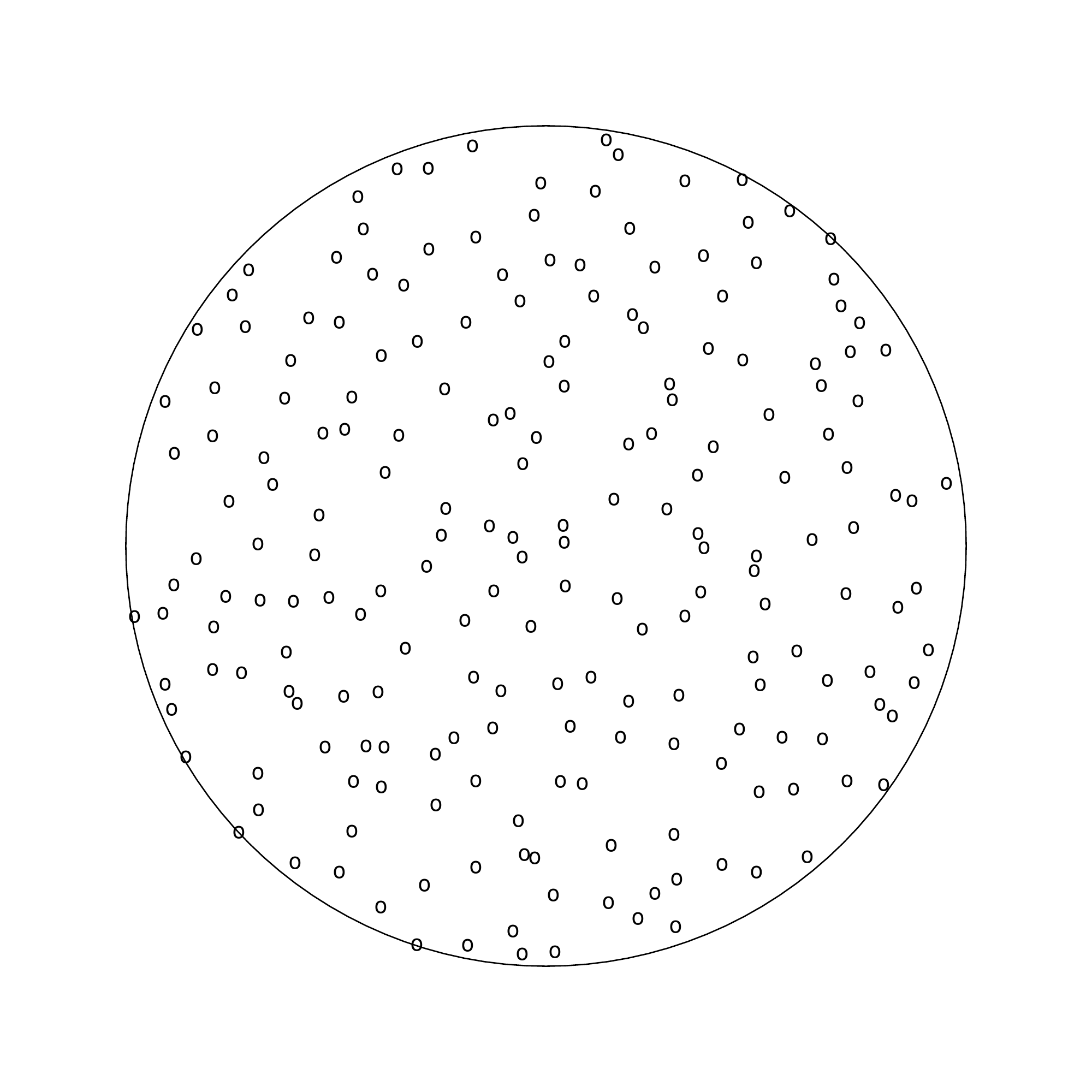}
  \end{subfigure}
  \vskip-5mm
  \begin{subfigure}[b]{0.5\textwidth}
    \includegraphics[width=\textwidth]{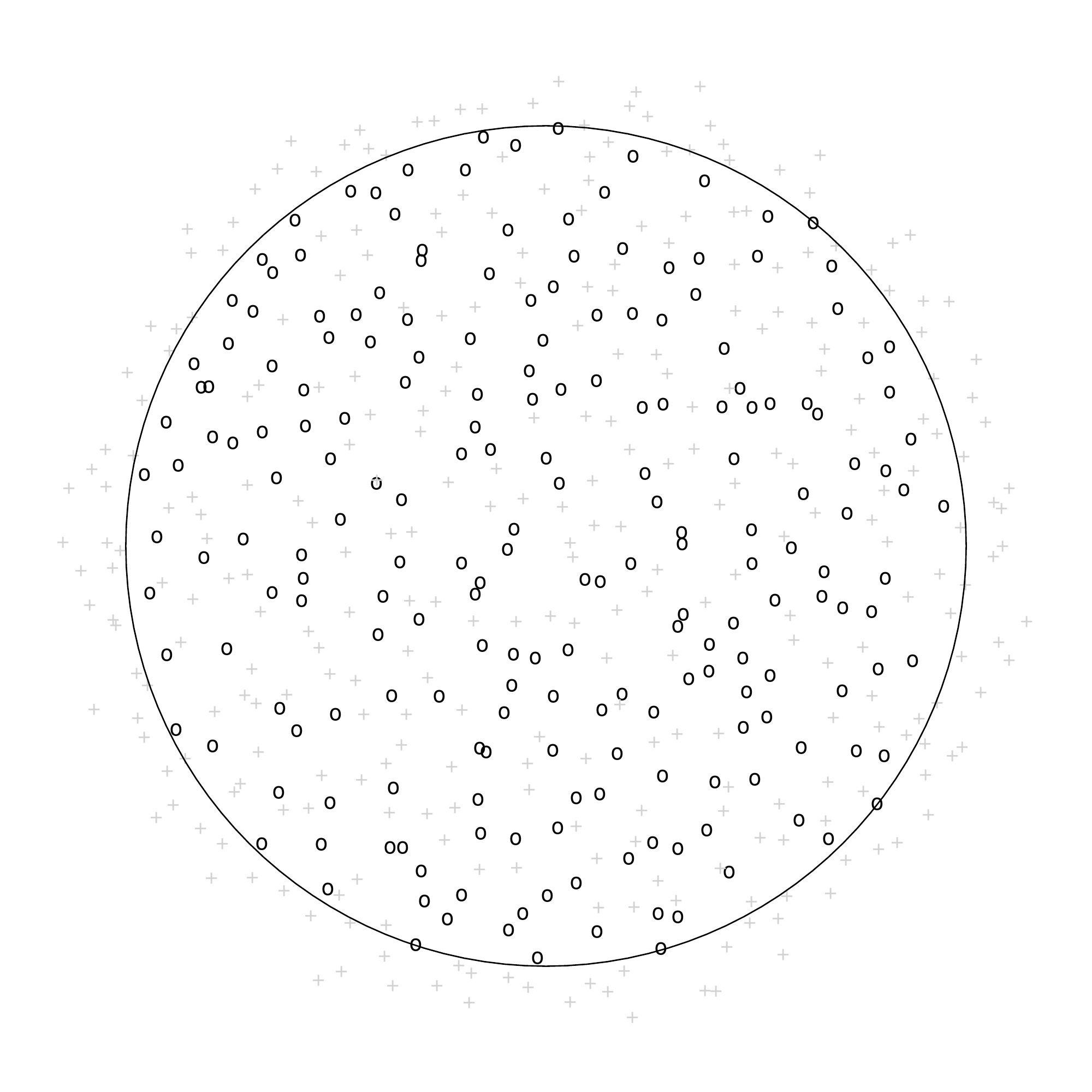}
  \end{subfigure}
  \begin{subfigure}[b]{0.5\textwidth}
    \includegraphics[width=\textwidth]{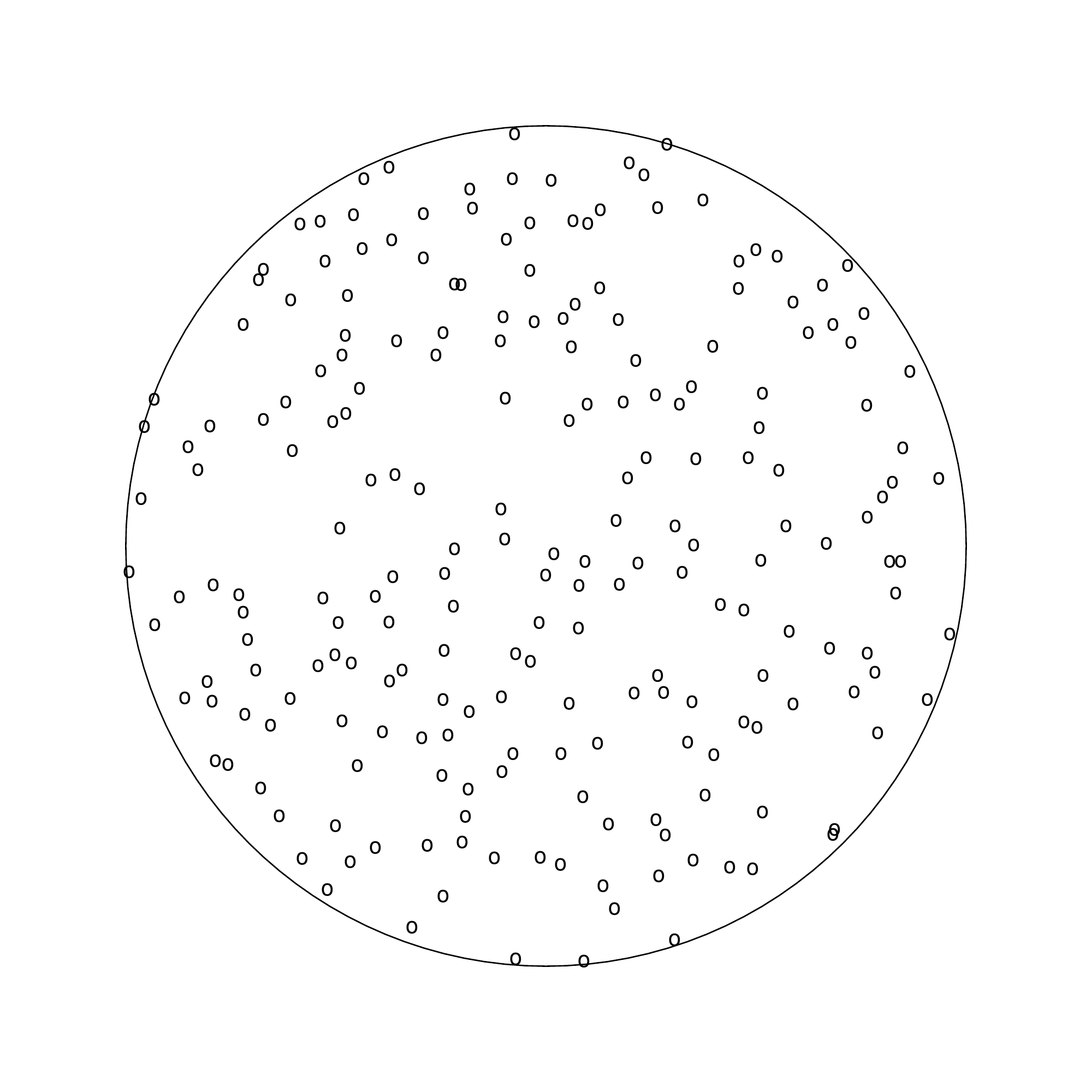}
  \end{subfigure}
  \caption{
      Realisations of two $\beta$-Ginibre processes generated with Algorithms~\ref{alg:ginibre_eigen} (left) and \ref{alg:ginibre_spectral} (right).
      In all cases $\rho=200$.
      In the top row $\beta = \beta_{\max}$ and in the bottom row $\beta = \beta_{\max}/2$.
      For Algorithm~\ref{alg:ginibre_eigen} on the left the gray points represent the points deleted by the algorithm.
  }
  \label{fig:Ginibre}
\end{figure}

In order to compare Algorithms~\ref{alg:ginibre_eigen} and \ref{alg:ginibre_spectral}, we have
performed several simulations of the $\beta$-Ginibre process, for different values of the
parameters $\rho$ and $\beta$,  on the ball $\B_R$ of area 1 where $R=1/\sqrt\pi$, so that the expected
number of points coincide with the intensity $\rho$.
An example  using each algorithm is shown in Figure~\ref{fig:Ginibre}, where the gray points on the
left correspond to the deleted points in the process of Algorithm~\ref{alg:ginibre_eigen}, while the black points represent the final simulated point pattern.

Table~\ref{tab:ginibre} displays the average computation time (over 500 replications) of each algorithm.
The picture is quite clear: the spectral algorithm \ref{alg:ginibre_spectral} is more efficient to generate	a moderate number of points, while Algorithm~\ref{alg:ginibre_eigen} becomes preferable for a high number of points.
On the other hand, the effect of $\beta$ is mild for the spectral algorithm, whereas it has a strong impact on the efficiency of Algorithm~\ref{alg:ginibre_eigen} (the smaller $\beta$ is, the less efficient it is), which is consistent with the fact that when $\beta$ is small, many points are generated to be afterwards deleted by the thinning step of Algorithm~\ref{alg:ginibre_eigen}, slowing down the process.

\begin{table}[ht]
\centering
\begin{tabular}{r|llll}
  \hline
Algo.~\ref{alg:ginibre_eigen}/\ref{alg:ginibre_spectral} & $\rho=100$ & $\rho=200$ & $\rho=400$ & $\rho=800$ \\
  \hline
$\beta=\beta_{\max}/3$ & 0.58/{\bf 0.09} & 2.99/{\bf 0.36} & 19.21/{\bf 6.63} & 130.14/{\bf 96.44} \\
  $\beta=\beta_{\max}/2$ & 0.17/{\bf 0.07} & 0.59/{\bf 0.27} & {\bf 3.11}/4.58 & {\bf 18.43}/68.83 \\
  $\beta=\beta_{\max}$ & 0.09/{\bf 0.07} & 0.28/{\bf 0.23} & {\bf 1.32}/3.43 & {\bf 6.45}/50.09 \\
   \hline
\end{tabular}
\caption{
  Average computation time (in seconds) for a single realization of a $\beta$-Ginibre process on $\B_R$, $R=1/\sqrt\pi$, with the given parameters, where $\beta_{\max}=1/(\rho\pi)$, using Algorithm~\ref{alg:ginibre_eigen} (left value) and Algorithm~\ref{alg:ginibre_spectral} (right value).
    The fastest method is boldfaced.}
  \label{tab:ginibre}
\end{table}

\section{Exact spectral expansion for other kernels}\label{sec:others}

This section aims at paving the way to (nearly) perfect simulation of the Gaussian-type DPP and the Bessel-type DPP, two widely used DPPs in spatial statistics.
In both cases we provide the explicit spectral representation of the kernel on the unit ball, which opens the way to perfect simulation by the spectral algorithm.
The numerical evaluation of the detailed spectral representations may pose some separate numerical challenges which are left for future research.

\subsection{Gaussian-type kernels}\label{sec:gaussian}

\subsubsection{Inhomogeneous kernel}\label{inhom gaussian}

We call the DPP with the following kernel an inhomogeneous Gaussian-type DPP:
\begin{equation}\label{gaussianbis}
 K_\sigma(x,y)= \rho  \, e^{-\|y-x\|^2/\alpha^2}\sqrt{p_\sigma(x)p_\sigma(y)},\quad x,y\in\R^d,
\end{equation}
where $\rho>0$ is the intensity parameter, $\alpha>0$ is the range parameter, and  $p_\sigma$ denotes the density of a standard normal distribution in $\R^d$ with standard deviation $\sigma>0$, i.e.
$$p_\sigma(x)=\frac{1}{(\sigma\sqrt{2\pi})^{d}} e^{-\|x\|^2/(2\sigma^2)},\quad x\in\R^d.$$
As verified in Appendix~\ref{sec:existenceKtilde}, this DPP exists if and only if 
\begin{equation}\label{existence inhomo}
2\rho^{1/d}\leq 1+\sqrt{1+8 \sigma^2/\alpha^2},
\end{equation}
or equivalently $2 \sigma^2/\alpha^2\geq \rho^{2/d} - \rho^{1/d}$.

Since the intensity is $K_\sigma(x,x)=\rho\, p_\sigma(x)$, $x\in \R^d$, this DPP is an inhomogeneous finite point process on $\R^d$ with expected cardinality $\rho$.
Accordingly, realised point patterns are concentrated near the origin.
Specifically, let $\chi^2_d(p)$ denote the quantile of the chi-squared distribution with $d$ degrees of freedom, then for any $0\leq\epsilon\leq 1$, the expected number of points in the centered ball with radius $\sigma\sqrt{\chi^2_d(1-\epsilon)}$ is $(1-\epsilon)\rho$.
For instance in dimension $d=2$, about $99\%$ of the points are in the centered ball with radius $3\sigma$.

A special property of the inhomogeneous Gaussian-type DPP is that the spectral representation of its kernel is perfectly known on $\R^d$, as explained below (see for instance \cite[Section~4.3.1]{williams2006}).
This decomposition specifically reads
\begin{equation}\label{spec gaussian} 
K_\sigma(x,y)=\sum_{j\in\mathbb N^d}\lambda_{j}\Phi_{j}(x)\Phi_{j}(y),\quad x,y\in\R^d,\end{equation}
where, setting $a=\sigma^{-2}/4$, $b=\alpha^{-2}$, $c=\sqrt{a^2+2ab}$, $A=a+b+c$ and $B=b/A$, for $j=(j_1,\dots,j_d)\in \mathbb N^d$ and $x=(x_1,\dots,x_d)\in\R^d$, 
\begin{equation}\label{eigen Ktilde}
\lambda_{j}=\rho \left(\frac{2a}A\right)^{d/2} B^{\sum_{k=1}^d j_k}
\end{equation}
and $\Phi_{j}(x)=\prod_{k=1}^d \varphi_{j_k}(x_{k})$ with
$$\varphi_k(x)=\left(\frac{\sqrt {2c}}{ 2^{k} k! \sqrt\pi}\right)^{1/2} e^{-c x^2}H_k(\sqrt{2c}\, x),$$
$H_k$ being the (physicists') Hermite polynomial of order $k$.
The spectral decomposition \eqref{spec gaussian} is a tensorial product and is deduced from the decomposition when $d=1$.
The latter is easily verified since when $d=1$, the identity $\int_\R K_\sigma (x,y) \varphi_k(x)dx=\rho \sqrt{2a/A}\,B^k \varphi_k(y)$ is a consequence of equation 7.374-8 in \cite{gradshteyn2014}, that is 
$$\int_\R e^{-(x-y)^2}H_k(tx)dx=\sqrt\pi (1- t^2)^{k/2}H_k(ty/\sqrt{1-t^2}),\quad |t|<1,y\in \R.$$
The orthonormality of the eigenfunctions is in turn a consequence of the relation $\int_\R H_{k_1}(x)H_{k_2}(x)e^{-x^2}dx =\sqrt\pi 2^k k!$ if $k_1=k_2$ and 0 otherwise.

Based on \eqref{spec gaussian}, the spectral algorithm is in theory feasible to simulate the inhomogeneous Gaussian-type DPP on $\R^d$.
Two approximations are however needed in practice.
The first one is the truncation of \eqref{spec gaussian} to  $|j|\leq \ell$, for some $\ell$ such that the expected number of generated points $\sum_{|j|\leq \ell} \lambda_j$ is sufficiently close to $\rho$. In fact, we know in theory the exact distribution of the index $\ell$ of the maximal non-null eigenvalue, and perfect simulation of $\ell$ could in principle be considered, see Appendix~D in \cite{lavancier_extended}. But this distribution is in practice impossible to generate and manual truncation is mandatory. 
This truncation allows us to generate the $(2\ell+1)^d$ Bernoulli variables $B_j$ that are needed in the first step of the spectral algorithm.
Then, given the $B_j$'s,  the simulation boils down to generating the projection DPP with kernel $K_{proj}(x,y)=\sum_{|j|\leq \ell} B_j \Phi_{j}(x)\Phi_{j}(y)$ using Algorithm~\ref{alg:sim}.
But in this algorithm, the simulation with respect to $p_i$ given by \eqref{eq:p_i} cannot easily be performed by Strategy 1 in Section~\ref{basic algo}
 because the simulation with respect to the unnormalised density  $K_{proj}(x,x)$ is not straightforward.
 Instead we need to apply Strategy 2, where the proposal in the rejection sampling is the uniform distribution.
 To do so, we need a second approximation which is the choice of a compact set $S$ that contains with high probability the generated point pattern.
 The natural choice for $S$ is the centered ball with radius $\sigma\sqrt{\chi^2_d(1-\epsilon)}$ for a prescribed error $\epsilon$.

\subsubsection{Homogeneous Gaussian-type kernel}
The homogeneous Gaussian-type kernel  is defined for any $x,y\in\R^d$ by
\begin{equation}\label{gaussian}
K(x,y)=\rho \, e^{-\|y-x\|^2/\alpha^2},\quad x,y\in\R^d.
\end{equation}
The DPP  with this kernel exists on $\R^d$ if and only if $\rho\alpha^d\pi^{d/2}\leq 1$ \cite{Lavancier}, and it is stationary with intensity $\rho$.

Note that the homogeneous Gaussian-type DPP is the limit in distribution, when $\sigma\to\infty$, of the inhomogeneous Gaussian-type DPP with intensity parameter $\rho\sigma^d(2\pi)^{d/2}$.
This comes from the fact that the convergence of the associated kernel holds  uniformly on all compact sets, which implies that the Laplace transforms of the two DPPs asymptotically coincide; see \cite[Proposition 3.10]{Shirai}.
Despite this property, the spectral representation of \eqref{gaussian} is not known, neither on $\R^d$ or on any subset.
For this reason, the standard approach to simulate the homogeneous Gaussian-type DPP is to resort to the spectral approximation by a Fourier series. 

We explain below an alternative simulation procedure, that is arguably much more accurate than the Fourier series approximation, although more time demanding.
The idea is to first generate an inhomogeneous  Gaussian-type DPP using the exact spectral representation \eqref{spec gaussian}, then apply an independent thinning procedure to get the final point pattern.

Assume that we wish to simulate the homogeneous Gaussian-type DPP with parameter $\rho$ and $\alpha$ on the unit ball $\B$ (the procedure extends straightforwardly to any compact set).
Let
$$\tilde\rho_\sigma = \rho \sup_{x\in \B} p_\sigma^{-1}(x)= \rho \sigma^d(2\pi)^{d/2} e^{1/(2\sigma^2)}.$$ 
The main steps of the simulation algorithm are the following:
\begin{enumerate}
\item Choose $\sigma_0=\underset{\sigma>0}{\mathrm{argmin}} \,\tilde{\rho}_\sigma$ subject to $\sigma^2\geq \alpha^2(\rho^{2/d}-\rho^{1/d})/2$.
\item Generate the inhomogeneous Gaussian-type DPP with intensity parameter $\tilde \rho_{\sigma_0}$, range parameter  $\alpha$ and standard deviation $\sigma_0$, on $\R^d$, using the spectral algorithm (see Section~\ref{inhom gaussian}).
\item Apply an independent thinning procedure with  retention probability 
$$q(x)=\frac{p_{\sigma_0}^{-1}(x)}{ (\sigma_0\sqrt{2\pi})^d e^{1/(2\sigma_0^2)}}  1_{\B}(x),\quad x\in\R^d.$$
\end{enumerate}

The first step aims at choosing the value of $\sigma$ that minimises the intensity of the inhomogeneous Gaussian-type DPP, thus optimising the computational cost, while ensuring the existence of the process through \eqref{existence inhomo}.
It is easily verified that the final point pattern is a DPP on $\R^d$ with kernel
$$\tilde\rho_{\sigma_0}  \, e^{-\|y-x\|^2/\alpha^2}\sqrt{p_{\sigma_0}(x)p_{\sigma_0}(y)} \times \sqrt{q(x)q(y)} = \rho \, e^{-\|y-x\|^2/\alpha^2} 1_{\B}(x)1_{\B}(y),$$
that is a homogeneous Gaussian-type DPP on $\B$.
This procedure provides a nearly perfect simulation algorithm of this DPP, up to the two mild approximations discussed in Section~\ref{inhom gaussian} used for the second step.
It can however be very time consuming, and we think that it is to be considered instead of the Fourier series approximation only if distributional accuracy is of main concern.

\subsection{Bessel-type kernel}\label{sec:bessel}

The Bessel-type kernel is defined for any $x,y\in\R^d$ by
\begin{equation}\label{bessel}
K(x,y)= \rho\,\Gamma\left(1+d/2\right) \frac{J_{d/2}\left(2\|x-y\|/\alpha\right)}{ (\|x-y\|/\alpha)^{d/2}},
\end{equation}
where $\rho>0$ is the intensity, $\alpha>0$ is the range parameter and $J_{d/2}$ is the Bessel function of the first kind.
This kernel defines a DPP in $\R^d$ whenever $\rho\,\alpha^d \pi^{d/2} \Gamma\left(1+d/2\right)  \leq 1$; see \cite{Biscio}.
Given $\rho>0$, the maximal possible value of  $\alpha$ corresponds to the most repulsive stationary DPP in $\R^d$ with intensity $\rho$, while $\alpha\to 0$ corresponds to the Poisson point process with intensity $\rho$.

As explained below, the spectral representation of this kernel is explicit on the unit ball $\B$ and involves the so-called generalized prolate spheroidal functions introduced and studied in \cite{slepian64}.
This representation opens the way to perfect simulation, by the spectral algorithm, of the Bessel-type DPP on any ball, and therefore on any bounded set (by first simulating on an outer ball and then taking the restriction to the desired set).

The generalized prolate spheroidal functions are the orthogonal eigenfunctions of the truncated Fourier transform, which means that given an arbitrary $c\in\R$, each such function $\psi$ satisfies for a certain $\gamma\in\C$
\begin{equation}\label{defpsi}\int_{\|y\|\leq 1} \psi(y) e^{c i x\cdot y} dy = \gamma \psi(x),\quad x\in\R^d.\end{equation}
The eigenvalues $\gamma$ and the eigenfunctions $\psi$ depend of course on $c$. 
These functions are proved to be real-valued, so iterating the latter equality and its transpose we obtain that
\begin{align*}
|\gamma|^2 \psi(x)= \int_{\|z\|\leq 1} \bar\gamma \psi(z) e^{c i x \cdot  z} dz &= \int_{\|z\|\leq 1}\left(\int_{\|y\|\leq 1} \psi(y) e^{-c i z \cdot  y} dy\right) e^{c i x \cdot  z} dz. 
\end{align*}
If we let $K_c$ denote the Fourier transform of the unit ball, i.e. (see \cite{grafakos2008classical})
$$K_c(x-y)=\int_{\|z\|\leq 1} e^{c i z\cdot (x-y)}dz=(2\pi)^{d/2} \frac{J_\frac{d}{2}\left(c\|x-y\|\right)}{(c\|x-y\|)^{d/2}},$$
this means that
$$\int_{\|y\|\leq 1} K_c(x-y) \psi(y)dy = |\gamma|^2 \psi(x).$$
By choosing $c=2/\alpha$, we deduce that $\psi$ is an eigenfunction of the Bessel-type kernel \eqref{bessel} associated with the eigenvalue $\rho\Gamma\left(1+d/2\right)|\gamma|^2/\pi^{d/2}$.

The generalized prolate spheroidal functions $\psi$ and the eigenvalues $\gamma$ have rather complicated expressions.
In the one dimensional case ($d=1$), these expressions are obtained in  \cite{slepian61} and detailed in \cite{moore04},  where the eigenvalues form a sequence $\gamma_n$ associated to the eigenfunctions $\psi_n$.
To make the connection with our setting, note that $\lambda_n(c)$ in \cite{moore04} corresponds to $c|\gamma_n|^2/(2\pi)$, and the associated eigenfunction $\psi_n(c,x)$ corresponds to $\psi_n(x)$.
Note finally that the norm of $\psi_n(c,x)$ on the unit ball  is $\sqrt {\lambda_n(c)}$ (see (9) in \cite{moore04}) so that in dimension $d=1$, the spectral expansion of the Bessel-type kernel on the unit ball  is
$$K(x,y)=\rho\alpha\frac \pi 2 \sum_{n\geq 0}\lambda_n^2(2/\alpha) \frac{\psi_n(2/\alpha,x)}{\sqrt{\lambda_n(2/\alpha)}}\frac{\psi_n(2/\alpha,y)}{\sqrt{\lambda_n(2/\alpha)}},\quad\forall x,y\in[-1,1],$$
using the notation of  \cite{moore04}.
Several algorithms are available to evaluate $\lambda_n$ and $\psi_n$; see for instance  \cite{moore04}, \cite{osipov2013}, \cite{Bremer2022} and the references therein.

In dimension $d\geq 2$, following \cite{slepian64}, the sequence of eigenvalues can be written $\{\gamma_{N,n}, N\geq 0,n\geq 0\}$, where the eigenvalues $\gamma_{0,n}$ are simple while the eigenvalues $\gamma_{N,n}$ for $N\geq 1$ have multiplicity $h(N,d)=(2N+d-2)(N+d-3)!/((d-2)!N!)$.
Accordingly, their associated eigenfunctions form  a sequence $\psi_{N,n,l}$ where $l=1,\dots,h(N,d)$.
All formulas are given in \cite{slepian64} and we do not reproduce them in the general case.
We only provide some details in Appendix~\ref{sec:prolate} for the case $d=2$; see also \cite{SHKOLNISKY07} for this particular case.  This results  in the following spectral representation, for any $x,y\in\B$,
\begin{multline}\label{spec bessel}
K(x,y)=2\pi\rho\alpha \sum_{n\geq 0} \lambda_{0,n}^2(2/\alpha) \psi_{0,n}(x)\psi_{0,n}(y) \\
+ 2\pi\rho\alpha\sum_{N\geq 1}\sum_{n\geq 0} \lambda_{N,n}^2(2/\alpha) (\psi_{N,n,1}(x)\psi_{N,n,1}(y) + \psi_{N,n,2}(x)\psi_{N,n,2}(y) ),
\end{multline}
where the specific form of $\lambda_{0,n}$, $\psi_{0,n}$,  $\lambda_{N,n}$, $\psi_{N,n,1}$ and $\psi_{N,n,2}$ are given in Appendix~\ref{sec:prolate}. 
Contrary to $d=1$, numerical algorithms to compute these eigenvalues and eigenfunctions in dimension $d=2$ are not mature yet, but they are the subject of an active current research and software is becoming available online \cite{SHKOLNISKY07,Greengard_18,Lederman_17}.
This will hopefully make the computation of the above spectral representation straightforward in the near future.

\appendix

\section{Conditional distributions}
\subsection{Palm distribution and Algorithms~\ref{alg:nospectral} and~\ref{alg:sim}}\label{appendix conditional}
Let $X$ be a DPP on $S$ with kernel $K$ and let $y_1,\dots,y_m\in S$.
Remember that the conditional distribution of $X\!\smallsetminus\!\{y_1,\dots,y_m\}$ given that $\{y_1,\dots,y_m\}\in X$ is formally called the reduced Palm distribution of $X$ given  $y_1,\dots,y_m$.
The intensity function of order $n$ of the reduced Palm distribution is 
\begin{equation}\label{palm intensity}
\rho_{y_1,\dots,y_m}^{(n)}(x_1,\dots,x_n)=\frac{\rho^{(n+m)}(x_1,\dots,x_n,y_1,\dots,y_m)}{\rho^{(m)}(y_1,\dots,y_m)}
\end{equation}
if $\rho^{(m)}(y_1,\dots,y_m)>0$, and 0 otherwise; see \cite{CL18}. 
Since $X$ is a DPP, we may simplify the expression of $\rho^{(n+m)}$ thanks to \eqref{defDPP-S} and the standard formula 
\begin{equation}\label{schur}
\det\begin{pmatrix} A & B \\ C & D \end{pmatrix} = \det(A) \det (D- CA^{-1} B),
\end{equation}
where $\det(A)=\rho^{(m)}(y_1,\dots,y_m)$, whereby we deduce that the reduced Palm distribution of $X$ given  $y_1,\dots,y_m$ is still a DPP with kernel 
\begin{equation}\label{conditional kernel}
K_{y_1,\dots,y_m}(x,y)=K(x,y)-{\bf k}^*_{m}(x) {\bf K}_{m}^{-1} {\bf k}_{m}(y),\quad x,y\in S,
\end{equation}
where ${\bf k}^*_{m}(x)=(K(x,y_1),\dots,K(x,y_m))$ and ${\bf K}_{m}=(K(y_j,y_l))_{1\leq j,l\leq m}$.

This result is not enough to simulate a general DPP, because it says nothing about its cardinality.
But if $X$ is a projection DPP on $S$, then we know its cardinality $n=\int_S K(x,x) dx$ and $\rho^{(n)}$ becomes the unnormalised density of the $n$-tuple set of points of $X$, with respect to the Lebesgue measure on $S^n$.
In this case, for any $i=n-1,\dots,1$, by \eqref{palm intensity} where $(m,n)$ is $(n-i,i)$, $\rho_{y_{i+1},\dots,y_n}^{(i)}$ is the unnormalised density of $X\!\smallsetminus\!\{y_{i+1},\dots,y_n\}$ given that $\{y_{i+1},\dots,y_n\}\in X$. 
We can thus apply a sequential conditional simulation: generate the first point $y_n$ of $X$, the marginal density of which is $K(x,x)/n$, then the second point $y_{n-1}$ given the first one, according to $K_{y_n}(x,x)/(n-1)$, then the third one given the first two points, according to $K_{y_n,y_{n-1}}(x,x)/(n-2)$, and so on.
This procedure leads to Algorithm~\ref{alg:nospectral}.

\medskip

To accelerate this algorithm, the expression of $K_{y_{i+1},\dots,y_n}(x,x)$ or equivalently of $ip_i(x)$ in \eqref{pi nospectral} can be rewritten to get \eqref{eq:p_i} in Algorithm~\ref{alg:sim}, provided we know the spectral representation of $K$ given by \eqref{spec_algo}. 
Then it is possible to calculate the matrix of the projection onto the orthogonal complement of $span\{\bv(X_n),\dots,\bv(X_{i+1})\}$ which is $P_i= I_n - {\bf V} ({\bf V}^*{\bf V})^{-1} {\bf V}^* = I_n - {\bf V} \,\overline{\bf K}_{i}^{-1}  {\bf V}^*$, where ${\bf V}$ is the $(n,n-i)$ matrix with entries $\Phi_i(X_j)$, $i=1,\dots,n$, $j=n,\dots,i+1$. 
Then, starting from \eqref{pi nospectral},  it is not difficult to verify that  $ip_i(x) = \|P_i \bv(x) \|^2$.
This means that  $ip_i$ may be obtained by successive orthogonalisation, which is exploited in Algorithm~\ref{alg:sim} (see \cite{Lavancier} for more details).

\subsection{Conditional distribution given an outside set}\label{outside set}

Let  $A$ and $A^c$ be a partition of $S$, i.e.\ $S=A\cup A^c$ and $A\cap A^c=\emptyset$.
If $X$ is a projection DPP on $S$ with cardinality $n$, then the density of $X\cap A$ given that $X\cap A^c=\{y_1,\dots,y_m\}$, $m < n$, is

\begin{align*}
f_{A|A^c}(X\cap A=\{x_1,\dots,x_{n-m}\} | X\cap A^c=&\{y_{1},\dots,y_{m}\}) = \\ &\frac{f_S(X=\{x_1,
  \dots, x_{n-m}, y_{1},\dots,y_{m}\})}{f_{A^c}(X\cap A^c=\{y_{1},\dots,y_{m}\})}
\end{align*}
if $\{x_1, \dots, x_{n-m}\}\subset A$ and 0 otherwise.
We have excluded the trivial case $m=n$ since $X\cap A$ is simply the empty set in this case.
Since $X$ is a projection DPP on $S$, the numerator above is (up to a multiplicative constant) the determinant of the $2\times 2$ block matrix 
$$\begin{pmatrix} [K(y_i,y_j)]_{1\leq i,j\leq m}  & [K(y_i,x_j)]_{1\leq i\leq m, 1\leq j \leq n-m} \\
[K(x_i,y_j)]_{1\leq i \leq n-m, 1\leq j\leq m}  & [K(x_i,x_j)]_{1\leq i,j\leq n-m}\end{pmatrix}.$$
By \eqref{schur} and using the same notation as in \eqref{conditional kernel}, we deduce that 
\begin{align*}
f_{A|A^c}(X\cap A=\{x_1, \dots, x_{n-m}\} | X\cap A^c= & \{y_{1},\dots,y_{m}\}) \propto \\ 
 & \det [K_{y_{1},\dots,y_{m}}(x_i,x_j)]_{1\leq i,j\leq n-m},
\end{align*}
where the constant of proportionality does not depend on $\{x_1,\dots,x_{n-m}\}$. 
This proves that given $X\cap A^c=\{y_1,\dots,y_m\}$, $X\cap A$ is a projection DPP with cardinality $n-m$ and kernel $K_{y_{1},\dots,y_{m}}(x,y)\1_A(x)\1_A(y)$.

\subsection{How to choose a projection DPP for conditional simulation?}\label{choice proj dpp}

We start by a simple setting.
Assume that we observe $m$ points of a point process $Y$ that is an independent thinning of a homogeneous point process $X$ on $S=[0,1]^d$ with constant thinning probability $\tau\in[0,1]$.
Assume moreover that we know the value of $\tau$.
Then a way to model $X$ as a projection DPP on $S$ that fits with the observed thinned point process $Y$ is as follows. 

Choose for $X$ a parametric stationary model of DPP with kernel $K(x,y)=\rho K_\alpha(x-y)$ with $K_\alpha(0)=1$, for which we know the Fourier transform $\varphi_\alpha$ of $K_\alpha$, for instance the Gaussian-type DPP model \citep{Lavancier}.
Here $\rho$ is the intensity of $X$ and $\alpha$ stands for the remaining parameters of the model, typically $\alpha$ is a univariate range parameter.  
Under this model, $Y$ is a DPP with kernel $K_{thin}(x,y)=\tau K(x,y)$ and so its intensity is $\tau\rho$ while its pair correlation function is $1-K_\alpha^2(x-y)$.
Consequently we can estimate from $Y$ the intensity of $X$ by $\hat\rho = m/\tau$ and the remaining parameter $\alpha$ by a contrast estimation method based on the pcf or the Ripley's $K$ function of $Y$ to get $\hat\alpha$ (see \cite{Lavancier, biscio:lavancier:17}).
Using the Fourier approximation of \cite{Lavancier}, we thus have $$K(x,y)\approx \hat\rho \sum_{k\geq 0} \varphi_{\hat\alpha}(k) e^{i k\cdot(x-y)},\quad x,y\in S.$$
Following the simulation procedure to generate a DPP with such kernel, a projection DPP kernel that can be used to approximate the distribution of $X$ is a realisation of 
$$K_{proj}(x,y)= \sum_{k\geq 0} B_k e^{i k\cdot(x-y)},$$
conditional on $\hat\rho$, $\hat\alpha$ and $\sum_{k\geq 0} B_k\geq m$, where the $B_k$'s are independent Bernoulli variables with respective rate $\hat\rho\varphi_{\hat\alpha}(k)$.

\bigskip
The above setting can be generalized to the case where the thinning probability is a known function $q(x)$, $x\in S$.
For instance we may have $q(x)=\tau a(x)/(\sup_{x\in S} a(x))$ where  $\tau\in[0,1]$ is known and $a(x)$ is an observed auxiliary variable.
Note that inpainting conditional simulation in $A$ given a point pattern in $A^c$  corresponds to the particular case where $q(x)=\1_{A^c}(x)$.
Then $Y$ becomes a DPP with kernel $K_{thin}(x,y)=\rho\sqrt{q(x)}K_\alpha(x-y)\sqrt{q(y)}$ and we can deduce $\hat\rho = m/(\int_S q(x)dx)$ and $\hat \alpha$ by minimum contrast estimation.
The approximation of $X$ by a projection DPP can then be carried out as above.

\bigskip

A further generalization is possible in the case where $X$ is inhomogeneous with intensity $\rho_\theta(x)$, where $\theta$ is an unknown parameter.
As before, we assume to observe $m$ points of $Y$, which is a thinning of $X$ with  thinning probability $q(x)=\tau a(x)/(\sup_{x\in S} a(x))$ where  $\tau\in[0,1]$ is known and $a(x)$ is an observed auxiliary variable. 
We can then start with a parametric DPP model for $X$ with kernel $K(x,y)=\sqrt{\rho_\theta(x)}K_\alpha(x-y)\sqrt{\rho_\theta(y)}$ from which we deduce that $Y$ is an inhomogeneous DPP with kernel $\sqrt{q(x)\rho_\theta(x)}K_\alpha(x-y)\sqrt{\rho_\theta(y)q(y)}$.
Based on the observation of $Y$, we may deduce the estimations  $\hat\theta$ and $\hat\alpha$ by a two-step estimation method as in \cite{poinas_sjs21}.
Remember that $X$ can be seen as a thinning of a homogeneous DPP $X^+$ with kernel $K^+(x,y)=\rho^+K_\alpha(x-y)$ and thinning probability $\rho_\theta(x)/\rho^+$, where $\rho^+$ is an upper bound of  $\rho_\theta(x)$ on $S$.
Following the same scheme as before, we may approximate $X^+$ by a projection DPP with random kernel
$$K_{proj}(x,y)= \sum_{k\geq 0} B_k e^{i k\cdot(x-y)},$$
conditional on $\hat\rho^+$, $\hat\alpha$ and $\sum_{k\geq 0} B_k\geq m$, where the $B_k$'s are independent Bernoulli variables with respective rate $\hat\rho^+\varphi_{\hat\alpha}(k)$.
A conditional simulation of $X^+$ given $Y$ is then possible.
To finally get a realisation of $X$ given $Y$, it suffices to thin the realisation of $X^+\!\smallsetminus\! Y$ with thinning probability $\rho_{\hat\theta}(x)/\hat\rho^+$.

\section{A bound on the conditional  probability density function for the Fourier basis}\label{sec:bound fourier}
From \cite{Lavancier}, we know that the conditional density $p_i$, given the $n-i$ first points $X_{i+1},\dots,X_n$ satisfies
\begin{equation}\label{bound-pi}
p_i(x)\leq    \frac{1}{i}  \min_{i+1\leq k\leq n} \left(K(x,x) -
  \frac{|K(x,X_k)|^2}{K(X_k,X_k)}\right).
\end{equation}
When using the Fourier basis on the unit square of $\R^d$, the projection kernel reads
\begin{equation*}
K(x,y)=\sum_{j\in J}
  \mathrm{e}^{2\mathrm{i}\pi j\cdot(x-y)}
\end{equation*}
where $J\subset \Z^d$ with cardinality $|J|=n$. 
The inequality \eqref{bound-pi}
becomes in this case 
\begin{equation*}
p_i(x)\leq    \frac{1}{i}  \left(n -
   \max_{i+1\leq k\leq n} \frac{|K(x,X_k)|^2}{n}\right).
\end{equation*}
But for any $x,y$ we have
\begin{align*}
|K(x,y)|^2 &= \left(\sum_{j\in J} \cos(2\pi j\cdot(x-y)) \right)^2 + \left(\sum_{j\in J} \sin(2\pi j\cdot(x-y)) \right)^2\\
&=\sum_{j,k\in J} \cos(2\pi j\cdot(x-y))\cos(2\pi k\cdot(x-y)) +  \sin(2\pi j\cdot(x-y))\sin(2\pi k\cdot(x-y))\\
&=\sum_{j,k\in J} \cos(2\pi (j-k)\cdot(x-y))\\
&=\sum_{p\geq 0} \frac{(-1)^p(2\pi)^{2p}}{(2p)!} \sum_{j,k\in J} [(j-k)\cdot(x-y)]^{2p}\\
&\geq n^2 - 2\pi^2  \sum_{j,k\in J} [(j-k)\cdot(x-y)]^{2}, 
\end{align*}
where the last inequality uses the fact that $|K(x,y)|^2$ is expanded into an alternating series.
Since  $\sum_{j,k\in J} [(j-k)\cdot(x-y)]^{2} =2 n  \sum_{j\in J} [j\cdot(x-y)]^{2} -  2 \left(\sum_{j\in J} j\cdot(x-y)\right)^{2}$, we obtain 
\begin{align*}
 i p_i(x) &\leq    n - 
   \max_{i+1\leq k\leq n} \left(n -  4  \pi^2 \sum_{j\in J} [j\cdot(x-X_k)]^{2} + \frac{4\pi^2} n  \left(\sum_{j\in J} j\cdot(x-X_k) \right)^{2} \right)_+.\\
   &=  \min_{i+1\leq k\leq n} \left(n,4  \pi^2 \sum_{j\in J} [j\cdot(x-X_k)]^{2} - \frac{4\pi^2} n  \left(\sum_{j\in J} j\cdot(x-X_k) \right)^{2} \right),
\end{align*}
which is \eqref{bound Fourier}.

\subsection{A new proposal in dimension $d=1$}\label{sec:fourierd1}

When $d=1$, we have $S=[0,1]$ and the inequality \eqref{bound Fourier} simplifies to give
$$ i p_i(x) \leq    \min_{i+1\leq k\leq n} \left(n, a^2 (x-X_k)^{2}  \right)$$
where $a^2= 4  \pi^2 (\sum_{j\in J} j^{2} -  (\sum_{j\in J} j)^{2}/n)$.

Note that $a^2 (x-X_k)^{2}\leq n$ iff $|x-X_k|\leq \sqrt n /a$.
The minimum  above thus simplifies greatly if all intervals $|x-X_k|\leq \sqrt n /a$, for $k=i+1,\dots,n$,  are disjoint. 
This motivates us to consider a thinning of $\{X_{i+1},\dots,X_n\}$ such that all points are at a distance greater than $2\sqrt n /a$ from each other and also at a distance larger than $\sqrt n /a$ from the borders 0 and 1.
This can be achieved by first removing the points too close to the borders, followed by a hardcore Matern~II thinning procedure \citep{baddeley:rubak:turner:15}.
We denote by $\{\tilde x_{1},\dots,\tilde x_p\}$ the retained points and we assume that they are ordered  ($\tilde x_{1}<\dots<\tilde x_p$).
We then have 
\begin{align*}
 i p_i(x) &\leq    \min_{1\leq k\leq p} \left(n, a^2 (x-\tilde x_k)^{2}  \right) \\
          & = \sum_{k=1}^p a^2 (x-\tilde x_k)^2 \1_{\B(\tilde x_k, \sqrt n/a)}(x) + n \1_{[0,1]\smallsetminus \bigcup \B(\tilde x_k, \sqrt n /a)}(x) \\
          & = \begin{cases}
            a^2 (x-\tilde x_k)^2 \text{ if } |x-\tilde x_k|<\sqrt{n}/a \text{ for some } k \in \{1,\dots,p\}\\
            n \text{ otherwise.}
            \end{cases}
 \end{align*}
Let us denote by $f(x)$ this upper-bound and by $F(x)$ its primitive. 
In order to generate a sample from $p_i(x)$ on $[0,1]$, we can use rejection sampling where the proposal has density $f(x)/F(1)$ and the rate of acceptance is $i/F(1)$.
Simulation from the proposal can be done by the inversion method: if $U\sim\mathcal U([0,1])$, $F^{-1}(F(1)U)$ is distributed from the proposal. 

To complete this procedure, it remains to provide the formulas for $F(1)$ and $F^{-1}(x)$. 
Note that for all $k$, 
$$\int_{\tilde x_k -\sqrt n /a}^{\tilde x_k +\sqrt n /a} a^2 (x-\tilde x_k)^2 dx = \frac{2}{3a} n ^{3/2}.$$
By construction of $\{\tilde x_{1},\dots,\tilde x_p\}$, we deduce that 
\begin{equation}\label{F1}
F(1)=p \frac{2}{3a} n ^{3/2} + n \left(1 - 2 p  \frac{\sqrt n} a\right)=n-\frac{4p}{3a} n^{3/2}.
\end{equation}
For $k=1,\dots,p$, we denote 
$y_k^-=F(\tilde x_k - \sqrt n/a)$, $y_k=F(\tilde x_k)$ and $y_k^+=F(\tilde x_k + \sqrt n/a)$ and we have
\begin{align*}
y_k^- & = (k-1) \frac{2}{3a} n ^{3/2}  + n \left(\tilde x_k -\frac{\sqrt n} a - (k-1) 2 \frac{\sqrt n} a\right) = n \tilde x_k - \frac{4k-1}{3a} n^{3/2},\\
y_k & =y_k^- + \frac{1}{3a} n ^{3/2}  = n \tilde x_k - \frac{4k-2}{3a} n^{3/2},\\
y_k^+& =y_k + \frac{1}{3a} n ^{3/2}  = n \tilde x_k - \frac{4k-3}{3a} n^{3/2}.
\end{align*}
For $ \tilde x_k - \sqrt n/a \leq x \leq \tilde x_k$, 
$$F(x)=y_k^-+ \int_{\tilde x_k - \sqrt n/a}^x a^2(t-\tilde x_k)^2 dt = n \tilde x_k - \frac{4k-2}{3a} n^{3/2} + \frac{a^2} 3 (x-\tilde x_k)^3.$$ 
The same results holds true when $ \tilde x_k \leq x \leq \tilde x_k + \sqrt n/a $.
We deduce that for $y_k^-\leq y\leq y_k^+$, 
$$F^{-1}(y) = \tilde x_k + \left(\frac{3}{a^2}\left(y-n\tilde x_k + (4k-2) n^{3/2}/a\right)\right)^{1/3}.$$
Finally, if $\tilde x_k + \sqrt n/a \leq x \leq  \tilde x_{k+1} - \sqrt n/a$, then $$F(x)=y_k^+ + n(x-(\tilde x_k + \sqrt n/a))=nx - \frac{4k}{3a} n^{3/2},$$
so that for $y_k^+ \leq y\leq y_{k+1}^-$, 
$$F^{-1}(y) = \frac 1n \left(y + \frac{4k}{3a} n^{3/2}\right).$$

With this new proposal, the acceptance rate in the rejection sampling for the last step of Algorithm~\ref{alg:sim}, i.e.\ for the simulation of the last point given the $n-1$ first ones, becomes $1/F(1)$ instead of $1/n$ for a uniform proposal.
For the most repulsive kernel  (i.e.\ $J=\{-\ell,\dots,\ell\}$ implying $n=2\ell +1$), we obtain  $a^2\leq \pi^2 n (n^2-1)/3$, so that this new rate can be $3$ times more effective, since by \eqref{F1} 
$$\frac{1/F(1)}{1/n}\geq \left(1-\frac{4p}{\pi \sqrt{3(n^2-1)}}\right)^{-1},$$
which is greater than 3 if  $p=n-1$ and $n>10$. 

\section{Existence of the inhomogeneous Gaussian-type DPP}\label{sec:existenceKtilde}
The existence is equivalent to $0\leq\lambda_i\leq 1$ for any $i\in \mathbb N^d$, where $\lambda_i$ is given by  \eqref{eigen Ktilde}.
By this formula, since $B\leq 1$, we have that for any $i$, $\lambda_i\leq\lambda_0$ and the condition is verified if only if $\rho  (2a/A)^{d/2}\leq 1$.
Since $A=a+b+\sqrt{a^2+2ab}$, this is equivalent to  $a(2\rho^{2/d} -1)-b\leq \sqrt{a^2+2ab}$, which is \eqref{existence inhomo}.

\section{Generalised prolate spheroidal wave functions in dimension $d=2$}\label{sec:prolate}

We provide in this appendix some details on the spectral representation \eqref{spec bessel} of the Bessel type kernel when $d=2$. As explained in Section~\ref{sec:bessel}, this amounts to describing the eigenvalues  and the associated generalized prolate spheroidal eigenfunctions in  \eqref{defpsi}, where $c=2/\alpha$. The following formulas come from \cite{slepian64} and  \cite{SHKOLNISKY07}.
When $d=2$, the sequence of eigenvalues in \eqref{defpsi} is $\{\gamma_{N,n}, N\geq 0,n\geq 0\}$ where the eigenvalues $\gamma_{0,n}$, $n\geq 0$, are simple, and  the eigenvalues $\gamma_{N,n}$ for $N\geq 1$ and $n\geq 0$ are associated to $h(N,2)=2$ eigenfunctions. 
Turning to polar coordinates by setting $x=(r\cos\theta,r\sin\theta)$ with $r\in[0,1]$ and $\theta\in[0,2\pi]$, the eigenfunctions read  for any  $n\geq 0$, 
$$\psi_{0,n}(x)=\frac 1 {\sqrt{2\pi}} R_{0,n}(r),\quad\forall x=(r\cos\theta,r\sin\theta)\in \B,$$ 
and for any $N\geq 1$ and $n\geq 0$, for all $x=(r\cos\theta,r\sin\theta)\in \B$,
\begin{align*}
\psi_{N,n,1}(x)=\frac 1 {\sqrt \pi} R_{N,n}(r)\cos (N\theta),\quad  \psi_{N,n,2}(x)=\frac 1 {\sqrt \pi} R_{N,n}(r)\sin (N\theta).
\end{align*}
In these expressions $R_{N,n}(r)$ is given in terms of a series of  Zernike polynomials:
\begin{equation}\label{zernike expansion}\sqrt r R_{N,n}(r)  = \sum_{k\geq 0} d_k^{N,n} T_{N,k}(r),\quad  \forall r\in [0,1].\end{equation}
In our notation the Zernike polynomials are defined for any $r\in [0,1]$ by 
\begin{equation}\label{def zernike}T_{N,k}(r) = \sqrt{2(2k+N+1)} r^{N+1/2} P_k^{N,0}(1-2r^2),\end{equation}
where $P_k^{N,0}$ are Jacobi polynomials, and they satisfy $<T_{N,k},T_{N,l}>=\delta_{k,l}$. (Note that the Zernike polynomials in  \cite{slepian64} are not normalised, unlike in \cite{SHKOLNISKY07} and in our case, explaining a slight difference in the definition). 
Concerning the coefficients $(d_k^{N,n})_{k\geq 0}$ in \eqref{zernike expansion}, they satisfy the recurrence relation 
$$b_{k,k-1}^N d_{k-1}^{N,n} + b_{k,k}^N d_k^{N,n} + b_{k,k+1}^N d_{k+1}^{N,n} = \chi^{N,n} d_k^{N,n},$$
where $\chi^{N,n}\in\R$ is an unknown quantity that does not depend on $k$ and where the coefficients $b_{i,j}^N$, the expression of which is given below,   are symmetric ($b_{i,j}^N=b_{j,i}^N$) and do not depend on $n$.
This means that given $N$, the (infinite) vector $(d_k^{N,n})_{k\geq 0}$ is the $n$-th eigenvector associated to the eigenvalue $\chi^{N,n}$ of the  symmetric tridiagonal (infinite) matrix $B^N$ with entries $b_{i,j}$:
$$B^N=\begin{pmatrix} 
b_{0,0}^{N} & b_{0,1}^{N} &  & &\\
b_{1,0}^N & b_{1,1}^N & b_{1,2}^N & &\\
& b_{2,1}^N & b_{2,2}^N & b_{2,3}^N & \\
&& \ddots & \ddots & \ddots & 
\end{pmatrix}
.$$
Here, for any $k\geq 0$, 
\begin{align*}
&b_{k,k-1}^N=c^2 \frac{k (k+N)}{(2k+N)\sqrt{2k+N+1}\sqrt{2k+N-1}},\\
&b_{k,k}^N=- \frac{c^2} 2 (1+ \frac{N^2}{(2k+N)(2k+N+2)}) - (2k+N+1/2) (2k+N+3/2),\\
&b_{k,k+1}^N=c^2 \frac{(k+1)(N+k+1)}{(2k+N+2)\sqrt{2k+N+1}\sqrt{2k+N+3}},
\end{align*}
and we recall that we have set $c=2/\alpha$.

For the numerical computation of \eqref{zernike expansion}, one needs for each $N$ to truncate the matrix $B^N$ at some size  $k_{\max}$, depending on the expected accuracy,  which allows to deduce all coefficients $d^{N,n}_k$ for $k\leq  k_{\max} -1$ and $n\leq  k_{\max} -1$.
The choice of  $k_{\max}$ is discussed in \cite{Greengard_18}. 
Note that in view of the orthonormal property of the Zernike polynomials, i.e. $<T_{N,k},T_{N,l}>=\delta_{k,l}$, the eigenfunctions $\psi_{0,n}$, $\psi_{N,n,1}$ and $\psi_{N,n,2}$ are orthonormal whenever the coefficients $(d_k^{N,n})_{k\geq 0}$ are normalised, i.e.\ $\sum_{k\geq 0} (d_k^{N,n})^2=1$, which is numerically achieved by normalising the eigenvectors of the (truncated) matrix $B^N$.

Concerning the eigenvalues in \eqref{spec bessel}, they are equal to 
$2\pi\rho\alpha\lambda_{N,n}^2(2/\alpha)$ where  $\lambda_{N,n}(c)$, $c=2/\alpha$,  satisfies (see (12) in  \cite{SHKOLNISKY07}):
\begin{equation}\label{eqlambda}
\lambda_{N,n}(c)\sqrt r R_{N,n}(r) = \int_0^1 J_N(crr')\sqrt{crr'} \sqrt {r'}   R_{N,n}(r') dr'.\end{equation}
Using \eqref{zernike expansion} and \eqref{def zernike}, we have for the left hand side
\begin{align}\label{equiv1}
\lambda_{N,n}(c)\frac{\sqrt r R_{N,n}(r)}{r^{N+1/2}}  &= \lambda_{N,n}(c) \sum_{k\geq 0} d_k^{N,n}\sqrt{2(2k+N+1)} P_k^{N,0}(1-2r^2) \nonumber \\
& \underset{r\to 0}{\sim} 
\lambda_{N,n}(c) \sum_{k\geq 0} d_k^{N,n}\sqrt{2(2k+N+1)} \binom{N+k}{k},
\end{align}
where the equivalence comes from the fact that for a Jacobi polynomial, $P_k^{N,0}(1)=\binom{N+k}{k}$.
For the right hand side in \eqref{eqlambda}, we have at $r=0$
$$\frac{J_N(crr')}{r^{N+1/2}}\sqrt{crr'} \sqrt {r'}   R_{N,n}(r')\sim \frac{(cr')^{N+1/2}}{2^N N!} \sqrt {r'}   R_{N,n}(r'),$$
so that using  \eqref{zernike expansion} again
$$\int_0^1 J_N(crr')\sqrt{crr'} \sqrt {r'}   R_{N,n}(r') dr'\sim  \frac{c^{N+1/2}}{2^N N!}\sum_{k\geq 0} d_k^{N,n} \int_0^1 {r'}^{N+1/2}T_{N,k}(r')dr'.$$
Since  $T_{N,0}(r)=\sqrt{2(N+1)}r^{N+1/2}$ and using the fact that $<T_{N,0},T_{N,k}>=\delta_{0,k}$, we obtain
\begin{equation}\label{equiv2}\int_0^1 J_N(crr')\sqrt{crr'} \sqrt {r'}   R_{N,n}(r') dr'\sim  \frac{c^{N+1/2}d_0^{N,n}}{2^N N!\sqrt{2(N+1)}}.\end{equation}
In view of \eqref{eqlambda}, the two right hand side expressions in \eqref{equiv1} and \eqref{equiv2} are equal, which gives 
\begin{equation}\label{deflambda}\lambda_{N,n}(c) = \frac{c^{N+1/2} d_0^{N,n}}{2^{N+1} N! \sqrt{N+1} \sum_{k\geq 0} d_k^{N,n} \sqrt{N+2k+1}\binom{N+k}{k}}.\end{equation}
For the numerical aspects, formula  \eqref{deflambda} may be implemented directly, or only for $\lambda_{N,0}$ and then a recursive relation can be used to get $\lambda_{N,n}$ for $n\geq 1$, as advised in dimension $d=1$ in \cite{osipov2013} through their relation (7.13), an approach generalised in $d\geq 2$ in \cite{Lederman_17,Greengard_18}.

\bibliographystyle{acm}
\bibliography{ref}
\end{document}